\documentclass[12pt,preprint2]{aastex}

\usepackage{multirow}
\usepackage{graphicx}
\usepackage{amsmath}
\usepackage{natbib}

\newcommand{\ec}{$\eta$~Car}
\newcommand{\degree}{\ensuremath{^\circ}}

\begin{document}

\title{SECULAR CHANGES IN ETA CARINAE'S WIND 1998--2011\altaffilmark{*,{\dagger},{\ddagger},\S,**}}

\author{Andrea Mehner\altaffilmark{1},
        Kris Davidson\altaffilmark{2},
        Roberta M.\ Humphreys\altaffilmark{2},
        Kazunori Ishibashi\altaffilmark{3},
        John C.\ Martin\altaffilmark{4},
        Mar\'{i}a Teresa Ruiz\altaffilmark{5}, and
        Frederick M. Walter\altaffilmark{6}   
	}

  \altaffiltext{*} {Based on observations made with the NASA/ESA Hubble Space Telescope, obtained from the Data Archive at the Space Telescope Science Institute, which is operated by the Association of Universities for Research in Astronomy, Inc., under NASA contract NAS 5-26555.}
  \altaffiltext{$\dagger$} {Based on observations obtained at the Gemini Observatory (acquired through the Gemini Science Archive), which is operated by the
Association of Universities for Research in Astronomy, Inc., under a cooperative agreement
with the NSF on behalf of the Gemini partnership: the National Science Foundation (United
States), the Science and Technology Facilities Council (United Kingdom), the
National Research Council (Canada), CONICYT (Chile), the Australian Research Council (Australia),
Minist\'{e}rio da Ci\^{e}ncia e Tecnologia (Brazil)
and Ministerio de Ciencia, Tecnolog\'{i}a e Innovaci\'{o}n Productiva (Argentina).}
  \altaffiltext{$\ddagger$} {Based on observations collected at the European Organisation for Astronomical Research in the Southern Hemisphere, Chile (obtained from the ESO Archive).}
  \altaffiltext{\S} {This paper includes data gathered with telescopes located at Las Campanas Observatory, Chile.}  
  \altaffiltext{**} {This paper includes data collected at Cerro Tololo Inter-American Observatory, Chile.}  
  
  \altaffiltext{1} {ESO, Alonso de Cordova 3107, Vitacura, Santiago de Chile, Chile}  
  \altaffiltext{2} {Minnesota Institute for Astrophysics, University of Minnesota, Minneapolis, MN 55455, USA}  
\altaffiltext{3} {Global COE, Division of Particle Physics and Astrophysics, Nagoya University, Nagoya 464-8602, Japan}  
\altaffiltext{4} {University of Illinois Springfield, Springfield, IL 62703, USA}
\altaffiltext{5} {Departamento de Astronom\'{i}a, Universidad de Chile, Casilla 36-D, Santiago de Chile, Chile}  
\altaffiltext{6} {Department of Physics and Astronomy, Stony Brook University, Stony Brook, NY 11794-3800, USA}

\begin{abstract}
Stellar wind-emission features in the spectrum of eta Carinae have decreased by factors of 1.5--3 relative to the continuum within the last 10 years. We investigate a large data set from several instruments (STIS, GMOS, UVES) obtained between 1998 and 2011 and we analyze the progression of spectral changes in the direct view of the star, in the reflected polar-on spectra at FOS4, and at the Weigelt knots. We find that the spectral changes occurred gradually on a time scale of about 10 years and that they are dependent on the viewing angle. The line strengths declined most in our direct view of the star. About a decade ago, broad stellar wind-emission features were much stronger in our line-of-sight view of the star than at FOS4. After the 2009 event, the wind-emission line strengths are now very similar at both locations.
High-excitation \ion{He}{1} and \ion{N}{2} absorption lines in direct view of the star strengthened gradually. The terminal velocity of Balmer P Cyg absorption lines now appears to be less latitude-dependent and the absorption strength may have weakened at FOS4.  
Latitude-dependent alterations in the mass-loss rate and the ionization structure of eta Carinae's wind are likely explanations for the observed spectral changes.

\end{abstract}


\section{INTRODUCTION}  
\label{sec:intro}

Eta Carinae, one of the most massive and most luminous stars in our Galaxy, is famous for its Great Eruption about 170 years ago.
Its recovery has been unsteady with unexplained photometric and spectral changes in the 1890s and 1940s (\citealt{2008AJ....135.1249H}, and references therein). The spectral changes described in this paper may represent another rapid step in \ec's recovery from its Great Eruption.

Eta Car has a complex spectroscopic cycle, most likely regulated by a companion star in an eccentric orbit (\citealt{1997NewA....2..107D}, and many references in \citealt{2005ASPC..332.....H} and \citealt{2012eta}).  
So-called spectroscopic events occur every 5.54 years since 1948 \citep{2001MNRAS.322..741F,1996ApJ...460L..49D,2008MNRAS.384.1649D}. The events are characterized by drastic changes in \ec's spectrum and photometry, e.g., high-excitation emission lines disappear for a few months (e.g., \citealt{1953ApJ...118..234G,1984A&A...137...79Z}) and light curves at all wavelength regions show significant variations (e.g., \citealt{1994MNRAS.270..364W,1997Natur.390..587C,2001MNRAS.322..741F,2006JAD....12....3V,2009A&A...493.1093F,2004AJ....127.2352M}).

In a previous paper \citep{2010ApJ...717L..22M} we compared spectra
at corresponding phases of successive spectroscopic cycles
and found dramatic changes in observations after the 2009 event.\footnote{
We define ``phase'' by $P = 2023.0$
days and $t_0 =$ MJD 50814.0 = J1998.00, consistent with the
Eta Carinae Treasury Program Archive (http://etacar.umn.edu/). Phases 0.00, 1.00, and 2.00 mark the 1998.0, 2003.5, and 2009.0
spectroscopic events.        
    }    
    Major stellar-wind emission features in the spectrum of \ec\ had decreased by factors of order 2 relative to the continuum within 10 years and helium P Cyg absorption had become stronger.
   Most of the broad emission lines in \ec's spectrum originate in the primary star's wind, see
   many papers and refs.\ in \citet{2005ASPC..332.....H}, and    
the simplest explanation for the observed spectral changes is a decrease in \ec's wind density,   by a factor of 2 or more.
The early exit from \ec's 2009 X-ray minimum and the observed decrease of the 2--10 keV photons over the last two cycles are consistent with this interpretation \citep{2009ApJ...701L..59K,2010ApJ...725.1528C,2011ApJ...740...80M}.

In this paper we analyze spectra obtained between 1998 and 2011 with several instruments to investigate in detail spectral changes in \ec's wind.  We are not concerned here with the temporary spectral changes observed during the events -- the spectral changes discussed are of secular nature. In \citet{2010ApJ...717L..22M} 
    we noted only a few examples;  here we explore a wider range of 
    effects, and whether or not they have developed gradually as opposed 
    to sporadically. 
Section {\ref{sec:obs}} describes the observations. In Section {\ref{sec:change}} we confirm the observations made by \citet{2010ApJ...717L..22M} and show that the broad stellar wind features were still weak in {\it HST\/} STIS data obtained several months after our initial discovery in 2010 March data. The temporal progression of spectral changes and the dependence on the viewing direction is discussed. High-excitation emission and continuum from the nearby Weigelt knots, which are thought to be photoionized by a hot companion star, reveal additional information. In Section {\ref{sec:massloss}} we discuss the implications of these observations and estimate the decrease in mass-loss rate over the last 10 years. In Section {\ref{sec:discussion}} we give a short summary.

\section{OBSERVATIONS AND DATA REDUCTION}
\label{sec:obs}

To investigate the long-term recovery of \ec\ from its Great Eruption, we need quantitative spectra with
consistent instrument characteristics, sampled over several
years.   Unfortunately, no suitable data set exists prior to the {\it Hubble Space Telescope\/}
({\it HST\/}) observations.  
{\it HST\/}  Space Telescope Imaging Spectrograph (STIS) observations in 1998--2004 and then again in 2009--2010 provide a consistent data set over a long time base-line. However, the STIS instrument was not available in 2004--2009, and  the position FOS4 in the southeast (SE) lobe
of the Homunculus, 4\farcs5 from the star, which shows the reflected pole-on spectrum was rarely observed with STIS.
We therefore supplemented the STIS observations with ground-based data from the
{\it Very Large Telescope\/} Ultraviolet and Visual Echelle Spectrograph ({\it VLT\/} UVES), the {\it Gemini-South\/} Gemini Multi-Object Spectrograph ({\it Gemini\/} GMOS), the {\it Magellan II\/} Magellan Inamori Kyocera Echelle ({\it Magellan II\/} MIKE), the {\it Ir\'{e}n\'{e}e du Pont\/} Boller \& Chivens Spectrograph ({\it Ir\'{e}n\'{e}e du Pont\/} B\&C),  and the {\it 1.5 m Cerro Tololo Interamerican Observatory\/} Ritchey-Chr\'{e}tien Spectrograph  ({\it 1.5 m CTIO\/} RC).    

{\it HST\/} STIS/CCD spectra obtained with the  
52{\arcsec}$\times$0\farcs1 slit in combination with the G230MB, G430M,
and G750M gratings covered the wavelength region from $\lambda\lambda$1700--10,000 \AA\ with spectral resolution $R \sim$ 5000--10,000.
The  observations include  a variety of slit positions and orientations covering the entire Homunculus nebula,
with a concentration at position angles $302\degree$ and $332\degree$ where the star
and the nearby ejecta, called Weigelt knots B, C, and D, fall within the slit.
The STIS data were reduced with improved reduction techniques that were developed for the
Eta Carinae HST Treasury Program \citep{2006hstc.conf..247D}.\footnote{   The reduced
{\it HST\/} STIS/CCD data can be downloaded from the Eta Carinae Treasury Project public
archive at http://etacar.umn.edu/. The reduction includes several improvements
over the normal STScI pipeline and standard CALSTIS reduction. Detailed information on the reduction procedures can be found on the website.}
We extracted one-dimensional spectra with a sampling width of 0\farcs13 using a mesa function \citep{2006ApJ...640..474M} at positions which were observed regularly; the central star and the Weigelt knots C and D.

{\it Gemini\/} GMOS spectra of the central object and FOS4 obtained in 2007--2010 provide valuable supplemental and  independent information.  
In most cases, we used the B1200 line grating at three tilt angles to
cover the spectrum from $\lambda\lambda$3700--7500 \AA.
A 0\farcs5-wide slit, oriented with a position angle of 160\degree,
was placed at different positions covering the star and
FOS4.  The resolving power was $R \sim$ 3000--6000.
The data reduction was done using the standard GMOS data reduction pipeline
in the Gemini IRAF package.
The spectra were extracted using a mesa function 11 by 7 pixels wide,
about 0\farcs8 by 0\farcs5.
The seeing was
roughly 0\farcs5--1\farcs5, so each GMOS spectrum discussed  
represents a region about 1{\arcsec} across. The spectra were rectified using a LOESS fit.\footnote{
For more information on the {\it Gemini\/} GMOS data and reduction procedures see the Technical Memo 14 at the Eta Carinae Treasury Project Website (http://etacar.umn.edu/treasury/techmemos/pdf/ tmemo014.pdf), \citet{2010AJ....139.2056M}, and \citet{2011ApJ...740...80M}.
}  

Unfortunately, the observations with {\it Gemini\/} GMOS do not cover an entire spectroscopic cycle.  Also, the important H$\alpha$ emission is so bright in \ec\ that it saturates the
detector pixels  even in the shortest available GMOS exposures centered on the star.
We therefore used observations obtained with the {\it VLT\/} UVES
instrument to examine in particular H$\alpha$ from 2002 to 2009.
The UVES observations are also extremely valuable because no other instrument covered the location at FOS4 consistently over such an extended time period. Eta Car was observed with UVES in the wavelength range from $\lambda\lambda$3000--8500 \AA\ using 0\farcs3 and 0\farcs4-wide slits. The slits were oriented with constant slit position angle of 160\degree\ and placed at two different positions covering the star and FOS4. The resolving power was $R \sim$ 80,000--110,000.
The data were reduced with the standard UVES pipeline available from ESO.\footnote{
      The reduced UVES observations can be downloaded from the Eta Carinae Treasury Project Website at http://etacar.umn.edu/.
      }  
Spectra were extracted using a mesa function 3 by 2 pixels wide,
about 0\farcs75 by 0\farcs5. The seeing was mostly between 0\farcs5 to 1\farcs5, with an average seeing of 0\farcs8.

The spatial resolution
of the {\it Gemini\/} GMOS and {\it VLT\/} UVES observations, limited by atmospheric seeing, is greatly inferior to
{\it HST\/} STIS spectra with spatial resolution better than 0\farcs2.
In ground-based observations, the inner ejecta are unavoidably
mixed with the spectrum of the star,  and  
include the Weigelt knots
0\farcs3 northwest of the star.  Fortunately, the slow-moving inner ejecta produce narrow emission lines
which are distinguishable from the broad stellar wind lines.
Typical widths are of the order of 20 and 400 km s$^{-1}$, respectively.  
At the wavelength region near $\lambda$4600 \AA, which is of interest in our analysis, the spectral resolution is about 40 km s$^{-1}$ for STIS,
roughly 75 km s$^{-1}$ for GMOS, and the UVES spectra have a spectral resolution better than 5 km s$^{-1}$.  
   Narrow lines are therefore more blurred in the GMOS data
while broad stellar wind features and their
P Cyg absorption components are well resolved by all three instruments.  

However, forbidden emission lines have an extended component at $\sim 0\farcs2$ from the central source \citep{2006ApJ...642.1098H}. The blue emission with velocities of $-$200 to $-$400 km s$^{-1}$ is located elongated along the NE-SW axis southwards of the central source \citep{2010ApJ...710..729M,2011ApJ...743L...3G}. The redshifted emission with velocities  of $+$100 to $+$200 km s$^{-1}$ is more asymmetric and extends towards the north-northwest \citep{2011ApJ...743L...3G}. These components are excluded in narrow extractions of the star in STIS observations but not in ground-based data which sample the inner $\sim 1\arcsec$ region. The broad stellar wind features near $\lambda$4600 \AA, discussed in Section \ref{sec:change}, normally include several forbidden lines and it is therefore non-trivial to compare {\it HST\/} with ground-based observations, see Section \ref{sec:long-term}.

In 2010 June we obtained observations with {\it Magellan II\/} MIKE, covering a wavelength region between $\lambda\lambda$3200--10,000 \AA. A 1{\arcsec} slit was used which resulted in spectral resolutions $R \sim$ 22,000--28,000, or about 10 km s$^{-1}$ near $\lambda$4600 \AA. The data were reduced with standard IRAF tasks and one-dimensional spectra corresponding to about 1{\arcsec} on the sky were extracted.

In 2011 February, June, and December we also obtained spectra of \ec\ and the FOS4 position with the B\&C spectrograph at the {\it Ir\'{e}n\'{e}e du Pont\/} telescope at Las Campanas Observatory. A 1\arcsec\ slit was used with the 1200/4000 grating centered at $\lambda$4500 \AA\ and the 1200/5000 grating centered at $\lambda$6000 \AA, covering the wavelength range $\lambda\lambda$3700--6700 \AA. The spectral resolution was $R \sim$ 2000--4000, or about 100 km s$^{-1}$ at $\lambda$4600 \AA. The seeing varied between 1--2\arcsec. The data were reduced using standard IRAF tasks and spectra were extracted using a mesa function with peak width of 2 pixels and base width of 4 pixels, corresponding to 1\farcs4 and 2\farcs8.

We obtained low resolution spectra with the RC spectrograph on the SMARTS  {\it 1.5 m CTIO\/} telescope in 2004--2012. A 2\arcsec\ slit and grating \#47 was used to  cover the wavelength range $\lambda\lambda$5650--6970 \AA\ with spectral resolution $R \sim$ 2000.  A 2\arcsec\ slit and grating \#26 covered the wavelength range $\lambda\lambda$3660--5440 \AA\ with spectral resolution $R \sim$ 1100. The data were reduced using standard procedures. Spectra are extracted by fitting a Gaussian plus a linear
background at each column and represent a region of $\sim 2\arcsec$ on the sky.

We also used {\it HST\/} STIS/MAMA observations of the central star with grating E140M and slit width of 0\farcs2, obtained between 2000 March and 2004 March, to investigate \ec's terminal wind velocity during the 2003.5 event using the \ion{Si}{2} $\lambda$1527 UV resonance line.\footnote{The reduced {\it HST\/} STIS/MAMA data can be downloaded from the Eta Carinae HST Treasury website at http://etacar.umn.edu/.} The spectral resolution is $R \sim 100,000$. We extracted spectra using a mesa function, corresponding to 0\farcs13.

Throughout this paper we quote vacuum wavelengths and heliocentric Doppler velocities.

\section{SPECTRAL CHANGES IN ETA CAR'S BROAD WIND-EMISSION FEATURES}
\label{sec:change}

\begin{figure}
\plotone{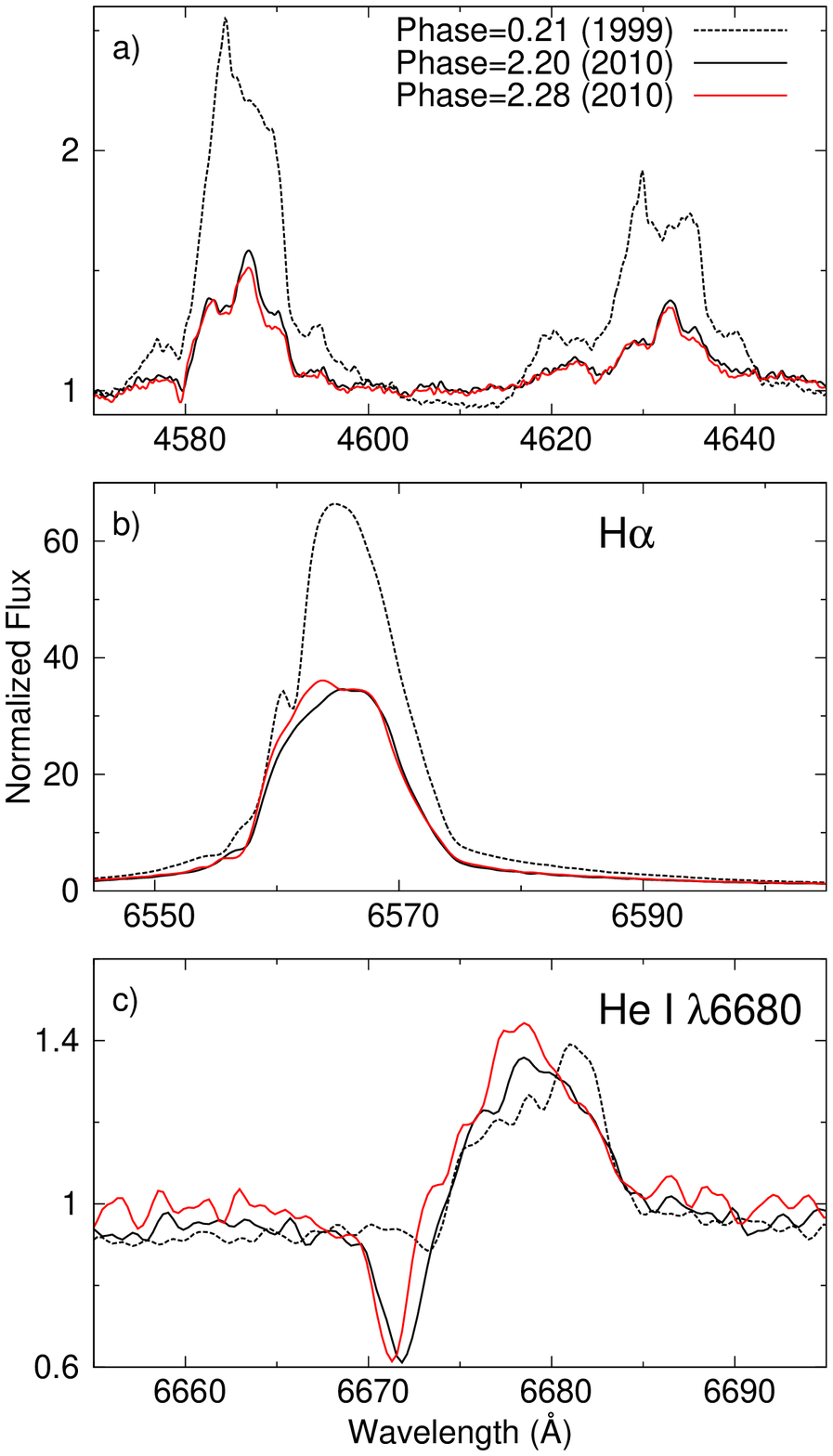}
\caption{{\it HST\/} STIS spectral tracings about 400 days after the 1998 and the 2009 events (phases 0.21 and 2.20) and about 570 days after the 2009 event (phase 2.28);
a) blends of \ion{Fe}{2}, [\ion{Fe}{2}], \ion{Cr}{2}, and [\ion{Cr}{2}] near $\lambda$4600 \AA, flux is normalized to unity at $\lambda$4740 \AA,
b) H$\alpha$, flux is normalized to unity at $\lambda$6630 \AA,
c) \ion{He}{1} $\lambda$6680, flux is normalized to unity at $\lambda$6630 \AA.  
The strengths of broad wind-emission features have not recovered in 2010 August (phase 2.28) observations. The external narrow absorption near $-144$ km s$^{-1}$ in the H$\alpha$ profile is still absent. \ion{He}{1} features shifted to bluer wavelengths and the \ion{He}{1} P Cyg absorption is still strong at phase 2.28. }  
\label{confirm}
\end{figure}

In \citet{2010ApJ...717L..22M} we reported dramatic changes in the broad wind-emission features of the central source in \ec. We compared spectra at corresponding phases of successive
cycles (phases 0.04 vs. 1.03, 1.12 vs. 2.10, and 0.21 vs. 2.20) and showed that the broad wind-emission features were considerably weaker in data obtained after the 2009 event, i.e., after phase 2.00, and that the \ion{He}{1} absorption had become unusually strong.
Observations with {\it HST\/} STIS obtained at phase 2.28 confirm these spectral changes, see Figure \ref{confirm}.  
The Figure shows spectral tracings of stellar wind features near $\lambda$4600 \AA, H$\alpha$, and \ion{He}{1} $\lambda$6680. In addition to the tracings at phases 0.21 (1999 February, $\sim 400$ days after the 1998 event) and 2.20 (2010 March 3, $\sim 400$ days after the 2009 event), which were already shown in \citet{2010ApJ...717L..22M}, the Figure includes observations at phase 2.28 (2010 August 20, $\sim 570$ days after the 2009 event). Between phases 2.21 and 2.28, the binary separation presumably increased by $\sim$ 14\% while the orbital longitude changed by about $\sim 5\degree$.
STIS observations in 2010 October (phase 2.31) did not cover H$\alpha$ and \ion{He}{1} $\lambda$6680 but sampled the broad wind features around $\lambda$4600 \AA.

Figure \ref{confirm}a shows broad \ion{Fe}{2}, [\ion{Fe}{2}], \ion{Cr}{2}, and [\ion{Cr}{2}] emission blends near $\lambda$4600 \AA\ which had decreased by a factor of 2--4 at phase 2.20 compared to phase 0.21. The strengths of the broad stellar wind features at phase 2.28 are comparable to the observations at phase 2.20.  STIS observations at phase 2.31 confirm further the secular nature of the weakened emission strengths, see Table \ref{tab:table1}.

Figure \ref{confirm}b confirms that the profile of H$\alpha$ is altered and weakened in the recent STIS data.
The narrow H$\alpha$ absorption near $-144$ km s$^{-1}$ seen in the tracing at phase 0.21  indicates unusual nebular physics far outside the wind \citep{2005A&A...435..183J}. This feature had weakened by 2007, reappeared during the 2009.0 event, but had practically vanished in March 2010 \citep{1984ApJ...285L..19R,1999ASPC..179..227D,2005AJ....129..900D,2010AJ....139.2056M, 2010AJ....139.1534R}. It is still absent in spectra obtained in 2011 December with {\it Ir\'{e}n\'{e}e du Pont\/} B\&C.
The H$\alpha$ profile at phase 2.28 is very similar to the one at phase 2.20  but shows an additional small blue emission feature on top. This component probably indicates the same or adjoining material as observed in the shifting \ion{He}{1} and \ion{N}{2} emission lines (\citealt{2011ApJ...737...70M}, compare also with Figure 1c). Note that well after the 2009 event, H$\alpha$ showed no signs of resuming what had once been its ``normal'' appearance.

High-excitation \ion{He}{1} emission did not weaken along with the features noted above, but the \ion{He}{1} P Cyg absorption greatly strengthened after the 2009 event. In observations at phase 2.28 the absorption is still strong, see Figure \ref{confirm}c. STIS observation of \ion{He}{1} $\lambda$4714 in 2010 October indicate that the helium absorption strengths may have even increased further compared to the 2010 August observations.
\ion{He}{1} emission and absorption lines shift to bluer wavelengths throughout \ec's spectroscopic cycle, compare tracings at phases 2.20 and 2.28 in the Figure (see also \citealt{2007ApJ...660..669N} and \citealt{2011ApJ...740...80M}).

Overall, we find that observations obtained in 2010 August (phase 2.28) compare well with observations obtained in 2010 March (phase 2.20); the wind did not change substantially in-between these observations. This is further confirmed by the analysis of the few spectral features observed in 2010 October (phase 2.31). The spectral change since 2004 is thus not simply a peculiarity or aftermath of the 2009 event, but probably represents a significant secular development in \ec's wind. We discuss the long-term nature of the spectral changes and their implications in the next sections.

\subsection{The Secular Character of the Spectral Changes}
\label{sec:long-term}

Spectral changes such as those found by \citet{2010ApJ...717L..22M} were expected in the long-term recovery of \ec\, but it was generally assumed that they would occur much more slowly.
The qualitative ground-based record from
1900 to 1990 showed no similar spectral changes in the broad wind-emission lines (excluding the events;  see many refs.\ in \citealt{2008AJ....135.1249H}). During 1991--2004, {\it HST\/} Faint Object Spectrograph (FOS) and STIS spectra showed no obvious  secular change
in $\eta$ Car's stellar wind spectrum.  
Figure 1a in \citet{2010ApJ...717L..22M},
illustrates the  similarity of the broad wind features in two successive cycles before 2004 at phases 0.04 and 1.03.
The 2009--2010 STIS data, however, revealed the weakest broad-line
spectrum ever seen in modern observations of \ec, relative
to the underlying continuum.
Low-excitation emission from the stellar wind became far
   less prominent on a time scale of only several years.  We suggest that a decrease of \ec's mass-loss rate is the most probable explanation \citep{2010ApJ...717L..22M}.  A precedent may have been the appearance of the high-excitation lines in the 1940s, probably also due to a change in the wind density \citep{2008AJ....135.1249H}.

To determine whether \ec's spectrum changed only after --  
     and as a result of -- the 2009 event, or if, alternatively,  
     the changes are of a more progressive nature, we investigated 
     spectra obtained since 1998 with several instruments.
The equivalent widths of two \ion{Fe}{2}/\ion{Cr}{2} blends near $\lambda$4600 \AA\ in data from 1998--2012 are listed in Table \ref{tab:table1}.
Ground-based observations, mainly with GMOS and UVES, fill in valuable data points during the years when STIS was unavailable, but they sample a wider region around the star and contain significant contributions from ejecta far outside
the stellar wind and from the broad extended emission component of forbidden lines, such as [\ion{Fe}{2}] and [\ion{Fe}{3}], mentioned in Section \ref{sec:obs}.  This results in very different equivalent width values for some broad wind features. Fortunately, several GMOS and UVES observations were obtained close to STIS observations, so we can correct for this effect as outlined below. 

For example, on 2009 June 30 the equivalent width of the $\lambda\lambda$4570--4600 \AA\ feature in STIS data was $EW$($\lambda\lambda${4570--4600,STIS}) = $3.42\pm0.27$ \AA. In UVES spectra on 2009 June 30 the equivalent width is $EW$($\lambda\lambda${4570--4600,STIS}) = $5.67\pm0.38$ \AA, a factor of 1.7 larger. Twenty four days later, on 2009 July 23, in GMOS spectra the equivalent width was $EW$($\lambda\lambda${4570--4600,GMOS}) = $6.50\pm0.31$ \AA, a factor of 1.9 larger.
Similarly, measurements for the blend at $\lambda\lambda$4614--4648 \AA\ are 1.6 times larger in UVES and 1.8 times larger in GMOS spectra when compared to STIS spectra, see Table \ref{tab:table1}. UVES and GMOS observations overlap during the years 2008 and 2009 and we consistently find somewhat smaller equivalent widths in UVES spectra compared to GMOS spectra, probably due to their better spatial resolution.
We use  the 2009 June STIS data set, which mapped the inner 1\arcsec\ region with slit offsets of 0\farcs1, to simulate a ground-based spectrum with a spatial sampling of 0\farcs65 by summing up the flux from the different slits. The equivalent widths from the simulated ground-based spectrum are; $EW$($\lambda\lambda${4570--4600,STIS,0.65"}) $\approx 6.2$ \AA\ and $EW$($\lambda\lambda${4614--4648,STIS,0.65"}) $\approx 5.3$ \AA. Those values agree well with the values obtained with UVES on the same day and the ones obtained about one month later with GMOS. The larger values found in ground-based data are therefore due to their larger spatial sampling.

\begin{figure}
\centering
\plotone{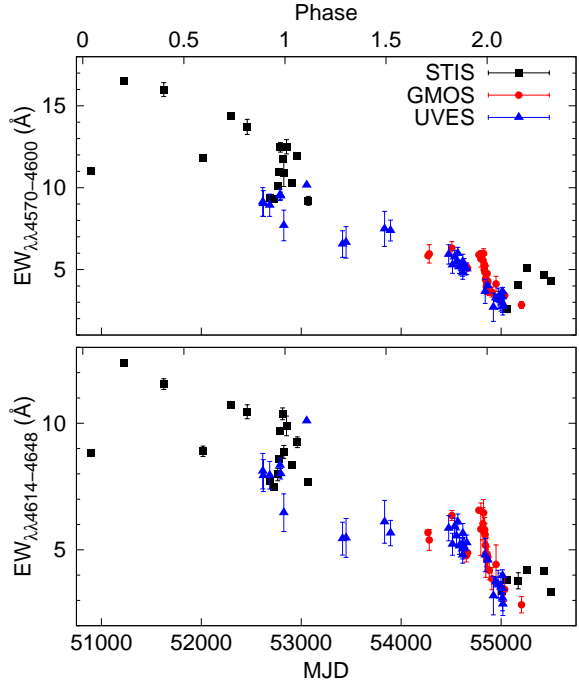}
\caption{Equivalent widths of \ion{Fe}{2}/\ion{Cr}{2} blends at $\lambda\lambda$4570--4600 \AA\ and $\lambda\lambda$4614--4648 \AA\  in {\it HST\/} STIS (black squares), {\it Gemini\/} GMOS (red circles), and {\it VLT\/} UVES (blue triangles) spectra in 1998--2010. Ground-based measurements were divided by 1.9 (GMOS, upper panel), 1.7 (UVES, upper panel), 1.8 (GMOS, lower panel), and 1.6 (UVES, lower panel) to account for the wider spatial sampling, see text. These broad stellar wind features show an almost linear decline over the last decade. \label{feII_long-term}}
\end{figure}

To compare the equivalent widths of the broad stellar wind features from different data sets, we adjust the values from the ground-based data using correction factors so that they are consistent with the values obtained from the STIS data in 2009. 
This approach may be questionable because 1) the inner and outer regions might not behave similar and 2) the 2009 data used to find the correction factors is very close to the 2009 event.
However, our method is justified because following this procedure we find that the UVES values in 2002 to 2004 then also overlap with the STIS values during those years.  
We therefore account for the  different spatial sampling of our ground-based data vs.\ the {\it HST\/} data  by applying correction factors, see Figure \ref{feII_long-term} (applied factors are given in the Figure caption). 

We find that the broad wind-emission features near $\lambda$4600 \AA\  decreased gradually by a factor of 2--3 over the last decade. Additional data sets, in particular the {\it 1.5 m CTIO\/} RC data, agree with this result, see Table \ref{tab:table1}. Neglecting
observations close to \ec's spectroscopic  events, near phases 1.0 and 2.0, when other factors dominate, the decline appears to be almost linear.

\begin{figure}
\plotone{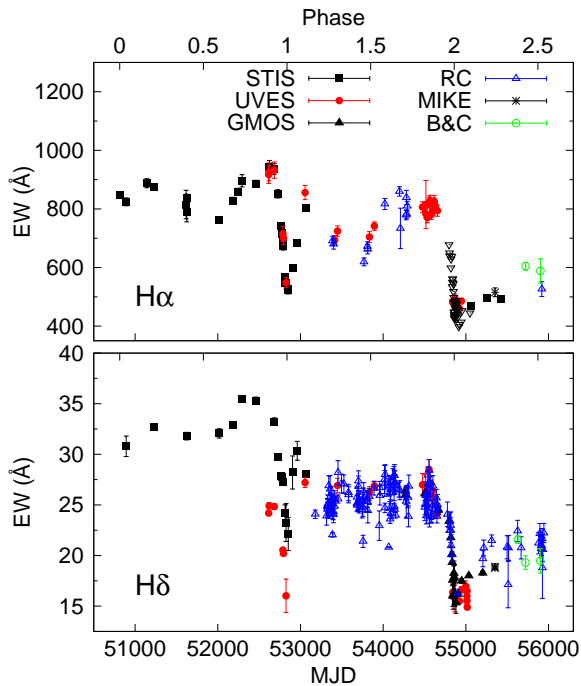}
\caption{Equivalent width of H$\alpha$ and H$\delta$ in 1998--2012. {\it HST\/} STIS (black squares) observations, which unfortunately were not available in 2004--2009, are supplemented by {\it VLT\/} UVES (red circles), {\it Gemini\/} GMOS (filled black triangles), and {\it Magellan II\/} MIKE (star) data. {\it Ir\'{e}n\'{e}e du Pont} B\&C (green circle) and {\it 1.5 m CTIO\/} RC data (blue triangles) are of lower quality. The open black triangles are from {\it 1.5 m CTIO} RC and Echelle observations and are retrieved from \citet{2010AJ....139.1534R}. The H$\alpha$ and H$\delta$ minima were deeper during the 2009 event compared to the 2003.5 event and the line strength did not recover afterwards. \label{Halpha_long-term}}
\end{figure}

We also monitored the H$\alpha$ and H$\delta$ equivalent widths in observations since 1998, see Table \ref{tab:table2} and Figure \ref{Halpha_long-term}.  {\it HST\/} STIS observations provide coverage over $\sim 12$ yr. In addition we analyzed data from the {\it VLT\/} UVES, {\it Gemini\/} GMOS, {\it Magellan II\/} MIKE,  {\it Ir\'{e}n\'{e}e du Pont\/} B\&C, and {\it 1.5 m CTIO\/} RC spectrographs. H$\alpha$ equivalent width measurements during the 2009 event with the {\it 1.5 m CTIO\/} RC and Echelle spectrographs retrieved from \citet{2010AJ....139.1534R} are also shown in the Figure. Unfortunately, H$\alpha$ could not be observed in direct view of the star with {\it Gemini\/} GMOS because the line is too bright even for the shortest allowed exposure times. No ``correction'' for different instruments as described above for the \ion{Fe}{2}/\ion{Cr}{2} blends is needed, since the total observed H$\alpha$ is dominated by 
     the stellar wind contribution even in ground-based data.

Figure \ref{Halpha_long-term} shows a subtle long-term trend to smaller H$\alpha$ and H$\delta$ emission line strengths by a factor of $\sim 1.5$ over the last decade, but the decline appears to be more pronounced after the 2009 event. Between 1998 and 2003 (phases 0--1) the strengths of Balmer emission remained within $\pm15 \%$  of their median value.
During the 2003.5 event, H$\alpha$ and H$\delta$ declined in $\sim 120$ days. H$\alpha$ then recovered in $\sim 200$ days and H$\delta$ faster in $\sim 120$ days. The 2009 event appeared, at first, to proceed similar to the previous event; the line strengths plummeted to a minimum in $\sim 120$ days. However, the minimum in 2009 was deeper than during the previous event and the emission did not recover to former strengths afterwards. A related note: Photometry at UV to visual wavelengths during the 2009 event also showed deeper minima in the light curves than in previous events \citep{2010NewA...15..108F,2011ApJ...740...80M}. \citet{2005AJ....129..900D} had already reported significant differences in the hydrogen line profiles between the 1998 and 2003.5 events; each event is distinct.  
Outside the events, if we view only the 
     data near phase $\sim 0.25$ of each cycle, then Figure  \ref{Halpha_long-term}  
     shows a linear trend somewhat like Figure \ref{feII_long-term}.
The gradual decrease of broad stellar wind-emission such as the \ion{Fe}{2}/\ion{Cr}{2} blends and hydrogen emission may represent a drop in \ec's mass-loss rate.

\citet{2012ApJ...746...73T} found no change in H$\delta$ line strength at phase $\sim 0.3$ in four consecutive cycles from 1994 to 2010. They argued that H$\delta$ is a better tracer of \ec's wind than, e.g., H$\alpha$ since it originates deep inside the primary's wind and is therefore less affected by the wind-wind collision region. Finding no changes in the H$\delta$ profiles they concluded that no changes occured in \ec's mass loss rate but that the changes reported by \citet{2010ApJ...717L..22M} were likely due to fluctuations in the wind-wind collision zone. However, Teodoro et al.\ only compared line profiles at one given phase and from two different data sets with inferior data quality than the data used in our analysis. Figure \ref{Halpha_long-term} shows that the trend described here is subtle and that individual measurements can fluctuate by up to $\sim 15 \%$ within days. To investigate the longterm trend a consistent measurement over the last decade as presented here is needed.

Hydrogen P Cyg absorption in our direct line of view is basically absent during \ec's normal state, but strong  P Cyg absorption develops for several months during the events \citep{2003ApJ...586..432S}, and was observed during the 2009 event \citep{2010AJ....139.1534R,2011ApJ...740...80M}. Unfortunately, we were unable to obtain unsaturated  H$\alpha$ profiles during the last event, but we did monitor H$\delta$ with GMOS. Before 2009 January only very weak H$\delta$ P Cyg absorption was observed. Strong absorption appeared suddenly between 2009 January 4 and 2009 January 9. In 2009 August STIS data the H$\alpha$ P Cyg absorption was absent but GMOS data still showed weak H$\delta$ P Cyg absorption in 2010 January.

Basic circumstances hamper the interpretation of \ec's Balmer 
   absorption lines.  Presumably they occur in zones where hydrogen is 
   mostly ionized, since the associated emission lines are very strong 
   and excitation to the $n = 2$ level is difficult in H$^0$ regions.  
   Therefore they depend on the ratio $n(H^{0},n=2)/n(H^{+})$, which is 
   small and sensitive to various effects that are hard to quantify  
   for a complex asymmetric wind.  Thus we cannot safely assume that 
   a Balmer absorption strength is well correlated with gas density, for 
   instance.  These difficulties have led to a major interpretational 
   disagreement between, e.g., \citet{2003ApJ...586..432S} and 
   \citet{2010AJ....139.1534R}, as mentioned below.

\begin{figure}
\centering
\plotone{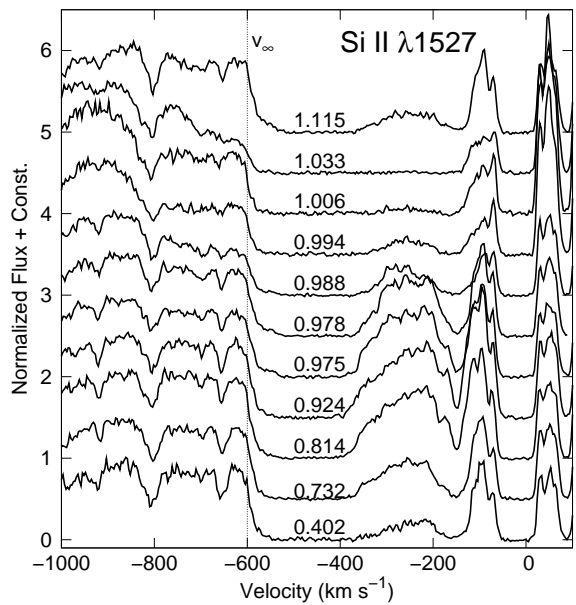}
\caption{\ion{Si}{2} $\lambda$1527 in {\it HST} STIS/MAMA observations in our direct line-of-sight to the central star. Phases are indicated next to each tracing and correspond to years 2000.23--2004.18. The terminal wind velocity  in UV resonance absorption lines during the 2003.5 event is constant. The differing emission strengths seen in these tracings are related to \ec's spectroscopic cycle. \label{UV_Si1527}}
\end{figure}

The terminal velocity of H$\delta$ P Cyg absorption was $v_{\infty} \sim - 550$ km s$^{-1}$ at all stellar latitudes in pre-event 2008 {\it Gemini} GMOS data  (see Section 5 in \citealt{2011ApJ...740...80M}). During the event, the terminal velocity of hydrogen absorption lines increased in our direct line-of-sight to about $v_{\infty} \sim - 900$ km s$^{-1}$. \citet{2003ApJ...586..432S} also found increasing terminal velocities of Balmer P Cyg absorption lines at moderate latitudes during the 1998 event. However, this does not necessarily imply that the velocity structure of \ec's wind changed.  UV resonance lines are better suited  to determine wind terminal velocities than Balmer lines. Unfortunately, no UV data were obtained during the 2009 event but {\it HST} STIS/MAMA covered \ec\ from 2000 to 2004. Figure \ref{UV_Si1527} compares \ion{Si}{2} $\lambda$1527 in spectra of the star in our direct line-of-sight showing a constant terminal velocity of \ec's equatorial wind of $v_{\infty} \sim -600$ km s$^{-1}$.\footnote{The constant terminal velocity of \ion{Si}{2} $\lambda$1527 may first be seen as an argument against a decreasing mass loss rate. However, the available UV data span only about 4 years from 2000 to 2004 (phases 0.4--1.1) and Figures \ref{feII_long-term} and \ref{Halpha_long-term} show no significant changes in the emission strengths of broad stellar wind features during this same time period. } Only at phase 1.033 a higher wind velocity might be possible, but the spectrum shortward of $-600$ km s$^{-1}$ can also be explained by the general weakening of emission lines,  visible in this same spectral region, during the event. The terminal velocity found in 1978 IUE data was comparable at $-600$ to $-700$ km s$^{-1}$ \citep{1979A&A....71L...9C}.
The appearance of hydrogen absorption lines in our line-of-sight to \ec\ and the increase of their terminal velocity may therefore result from changes in the ionization structure 
      of \ec's wind modulated by the secondary star's UV radiation
      \citep{2010AJ....139.1534R} or a wind cavity \citep{2011arXiv1111.2280M}, and not from a change in the mass-loss 
      structure as proposed by \citet{2003ApJ...586..432S}.

Helium emission and absorption processes in \ec's wind depend on the companion star and have
other special characteristics, see Section 6 of \citet{2008AJ....135.1249H}.  
  Similar to the case of a photoionized nebula, the amount of \ion{He}{1} 
  emission depends mainly on the hot companion star's helium-ionizing 
  photon output ($h{\nu} \gtrsim 25$ eV), with only weak dependences on 
  the location of the recombining He$^+$, gas density, and other details. 
    Therefore it is not surprising that the helium emission lines
    behave differently from the lower-excitation features.
The equivalent widths of \ion{He}{1}  emission lines remained constant from cycle to cycle. However, after the 2009 event, the \ion{He}{1} P Cyg {\it absorption\/} strength had greatly increased compared to previous cycles \citep{2010ApJ...717L..22M}.   \cite{2004IBVS.5492....1G} had already noted increasing \ion{He}{1} $\lambda$6680 P Cyg absorption from 1992 to 2003.
STIS observations since 1998 show that \ion{He}{1} absorption in spectra in direct view of the central source was very weak shortly after the 1998 event, but increased until 2003. During the 2003.5 event, the absorption vanished, but reappeared shortly after. GMOS observations, starting in 2007 about 600 days before the 2009 event, show that the absorption increased further. It then again disappeared during the 2009 event but became very strong by mid-2009. Overall, the \ion{He}{1} absorption strengths increased since 1998, only  interrupted by episodes close to the events when the absorption disappeared for a few months.
The same behavior is also observed for the \ion{N}{2} $\lambda\lambda$5668--5712 series, discussed in \citet{2011ApJ...737...70M}.

\begin{figure}
\centering
\plotone{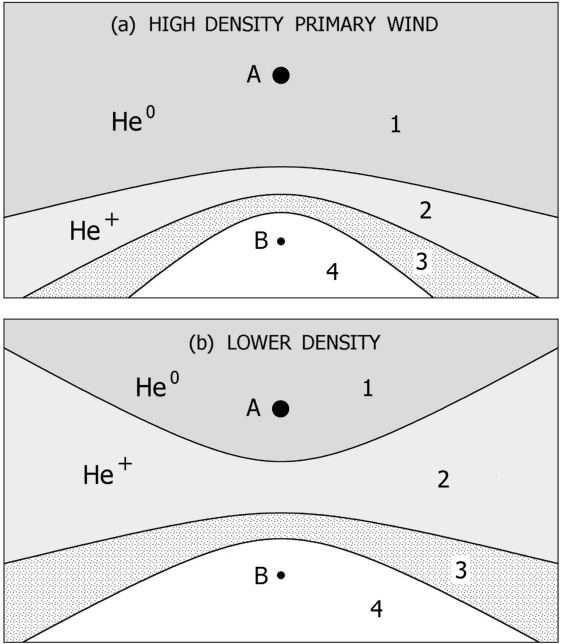}
\caption{Schematic helium ionization zones in $\eta$ Car's wind.  A and B 
   are the two stars.  Zones 1 and 2 occur in the undisturbed primary wind, 
   zone 3 is the colliding-wind shocked region, and zone 4 is the low-density
   secondary wind.  Observed \ion{He}{1} recombination emission arises 
   mainly in zone 2, where helium is photoionized by the hot secondary star B.  
   (Small condensations in zone 3 can also produce
     appreciable  \ion{He}{1} emission, but zone 4 is
     insufficiently dense.) The recent decrease of the primary wind may 
   have enlarged the geometrical extent of zone 2 as shown in the bottom 
   panel.  {\it Caveat:\/} This diagram is highly idealized;
   for example, the boundary between zones 2 and 3 is quite
   unstable, irregular and ill-defined. \label{ion_zone}}
\end{figure}

Changes in \ec's mass-loss rate help to explain these observations, because 
a lower wind density automatically implies larger photoionized zones.  
Since the observed \ion{He}{1} absorption lines arise from highly excited 
levels, they are indirect consequences of recombination in He$^+$ zones,  
not He$^0$ \citep{2006agna.book.....O};  and the He$^+$ gas is probably 
more extended than it was 10 years ago.  Two plausible locations have 
been suggested, as sketched in Figure \ref{ion_zone}.  
  \begin{enumerate} 
  \item { One is the shocked colliding-wind region, zone 3 in 
  the figure \citep{2008AJ....135.1249H,2008MNRAS.386.2330D}.  Most of the volume there has 
  He$^{++}$ at $T > 10^6$ K, but small cooled condensations also exist 
  (see below).  If they intercept most of the secondary star's 
  helium-ionizing photons, then they contain the relevant He$^+$ gas,   
  and zone 2 in Figure \ref{ion_zone} shrinks to negligible thickness.     
  In this case a change in the shocked zone, e.g., an increased opening 
  angle, may explain the increasing \ion{He}{1} P Cyg absorption 
  \citep{2010A&A...517A...9G}.  In our view, the most likely reason for 
  this to occur is a decrease in the primary wind outflow.  This would 
  move the shocked region closer to the primary star while broadening 
  its opening angle -- thus tending to increase the range of directions 
  where a line of sight intersects appreciable He$^+$. }  
  \item { On the other hand, as we explain below, the parameters strongly 
  suggest that many of the secondary star's ionizing photons pass between 
  the small shocked-and-cooled condensations, penetrate into the primary 
  wind,  and form zone 2 in Figure \ref{ion_zone}.   As the figure shows, 
  this region   becomes dramatically larger if the primary wind density 
  decreases by a factor of three.\footnote{  
     Figure \ref{ion_zone} is only a sketch and the parameters are poorly 
     known, but it is realistic in an order-of-magnitude sense.    
     The He$^+$ ionization fronts were estimated from Zanstra
     calculations for $r^{-2}$ density distributions at $T \gtrsim 10^4$ K
     \citep{2006agna.book.....O}.  Primary mass loss rates of roughly 
     $10^{-3}$ and $3 \times 10^{-4}$ $M_\odot$ yr$^{-1}$ were assumed, 
     with a secondary star having $L \approx 4 \times 10^5$ $L_\odot$ and
     $T_\textnormal{eff} \approx$ 40,000 K.  Extra ionization by the 
     primary star was included \citep{2008AJ....135.1249H}, and UV 
     absorption in shocked zone 3 was neglected.  In reality the 
     distinction between zones 2 and 3 is ill-defined on the spatial scale 
     shown here, because the primary shock is very unstable. }  
  The upper panel of the figure represents a dense wind, arguably like 
  \ec's state before 2004.  In that case, most geometric rays from
  the primary star do not intersect any He$^+$ gas.  With the orbit
  orientation favored by most authors (e.g., 
  \citealt{2008MNRAS.388L..39O,2009MNRAS.394.1758P,2012MNRAS.tmp.2292M}),
  our line of sight to the primary star would pass through the 
  quasi-hyperboloidal He$^+$ zone only for a limited time near conjunction, 
  3--11 months before periastron -- depending of course on the orbit 
  orientation and the shock-front opening angle.  At other times,
  there would be little or no He$^+$ along the line of sight.  (The 
  same statement applies to the shocked colliding-wind zone.)  
  The lower panel of Figure \ref{ion_zone}, by contrast, has a
  far broader He$^+$ zone because the wind is less dense by a factor
  of about 3.  It notionally represents the situation today.  In this
  case, our line of sight passes through He$^+$ during {\it most\/}
  of the orbit, except for two or three months before and after periastron.
  Therefore a decreased wind density improves the observability of 
  \ion{He}{1} absorption, while having little effect on the \ion{He}{1} 
  emission strengths as we noted earlier.  }  
  \end{enumerate}  
In principle, zones 2 and 3 in Figure \ref{ion_zone} may be of comparable 
importance for the \ion{He}{1} lines.  In both cases a decreased wind 
density appears to be consistent with the data.

Which of the above views is more accurate?  Unfortunately the ionization 
problem is extremely intricate within the shocked gas.   Consider, for 
example,  the primary-wind shock at a time when it is located 15 AU from 
the primary star.  For the sake of discussion, suppose the wind speed is 
500 km s$^{-1}$, the total mass loss rate is 
$3 \times 10^{-4}$ $M_\odot$ yr$^{-1}$, and ignore likely inhomogeneities 
in the wind.\footnote{
    An assumed 500 km s$^{-1}$ wind speed is merely conventional.   
    Judging from the bipolar structure of $\eta$ Car's ejecta, 
    the outflow may be considerably slower at equatorial latitudes.}   
Then an idealized adiabatic shock produces post-shock temperature 
$T \sim 4 \times 10^6$ K and electron density $n_e \sim 10^9$ cm$^{-3}$.  
But the cooling time is $t_c \sim 10^5$ s 
(Chapter 34 in \citealt{2011piim.book.....D}), much faster than the 
outflow escape time $t_{esc} \sim 4 \times 10^6$ s.  
Trapped X-rays may delay the cooling,  but not enough to alter the 
basic situation.  Therefore a naive one-dimensional shock model  
has a sheet of cooled gas with $T <$ 20,000 K.  This gas is much denser 
than the pre-shock wind, because pressure equilibrium applies in an 
approximate sense between the two shock fronts.  Consequently it  
would block practically all incident ionizing photons, so zone 2 in Figure 
\ref{ion_zone} would not exist.  This simple view is obviously unrealistic, 
though, because a number of well-known thermal, fluid, and radiation 
instabilities disrupt the sheet as rapidly as it forms.   Figure 7 in 
\citet{1992ApJ...386..265S} illustrates this 
phenomenon in a 2-dimensional model, and the case of $\eta$ Car is even 
more dramatic for two reasons:  3-dimensional geometry allows the 
development of small condensations, and the radiation pressure of 
ionizing radiation from the secondary star incites an additional, 
Rayleigh-Taylor-like instability.  Hence there is little doubt that 
rapid cooling forms a fine spray of blob-like or filament-like 
condensations.  Figure 7 in Stevens et al.\ hints that these may 
form streamers pointed toward the secondary star.  Meanwhile, hot 
shocked gas between the condensations ($T \gtrsim 10^6$ K) contains 
He$^{++}$ and is nearly transparent to ionizing UV radiation.  
Evidently the question at hand is:  {\it Do the many small condensations 
intercept most of the UV photons incident on the shock structure?\/}  
If they do, then \ion{He}{1} emission and absorption arises 
mainly within the colliding-wind zone;  but otherwise, zone 2 in our 
Figure \ref{ion_zone} is more important.

Let us attempt an order-of-magnitude estimate with the same parameters 
assumed above.  For simplicity we assume that each cooled condensation 
is a ``blob'' rather than a filament; if necessary a filament might be 
represented as a line of blobs.  The characteristic pre-cooling size 
scale for thermal instability is of the order of $w t_c \sim 0.15$ AU,  
where $w \sim 200$ km s$^{-1}$ is the adiabatic-shocked sound speed.  
Cooling rapidly shrinks this size scale to less than 0.03 AU, about 
1 percent of the colliding-wind region's overall size scale.  (The 
shrinkage factor is the cube root of the density-increase factor.)  
One expects roughly 300 blobs per AU$^3$ (i.e., one per 0.15 AU cube) 
in the shocked region which is about 3 AU thick.   Thus we expect a column 
density $N_b \sim 10^3$ blobs  per AU$^2$.  If the geometrical cross-section 
of each blob is $\sigma_b \sim (0.03 \; \mathrm{AU})^2  \sim 10^{-3}$ AU$^2$, 
then we find an ``equivalent optical depth'' $\tau_b = N_b \sigma_b \sim 1$, 
meaning that comparable numbers of photons either do or do not penetrate 
through the shocked region.   Most of the factors neglected here would 
tend to decrease $\tau_b$.  For instance, radiation pressure tends 
to either disrupt or ablate a blob on a time scale less than $t_{esc}$; 
and blobs may tend to be aligned with the direction to the secondary star,  
thereby increasing the transparency of regions between such filaments.     
In summary, the issue is left in doubt, because we can do only an 
order-of-magnitude assessment.  No computer codes applied to $\eta$ Car 
so far can realistically solve this problem, because a satisfactory 
model requires 
(1) 3-dimensional fluid dynamics with spatial resolution $\sim \, 10^3$,   
(2) realistic thermal and ionization microphysics including possible 
ablation,   
(3) realistic 3-dimensional radiative transfer for the ionizing photons, and 
(4) valid input parameters.
None of these can be omitted.  
This puzzle is so intricate that tempting approximations may lead to 
serious errors.  One interesting detail  is that each condensation 
may move semi-ballistically, being too small and dense to follow the 
general fluid flow;  while the ionizing-radiation pressure is not very 
much smaller than the thermal gas pressure.   A final remark on this 
sub-topic:  In view of the very strong instabilities of the primary-wind 
shock, exacerbated by inhomogeneities in the primary wind, the 
boundary between zones 2 and 3  in Figure \ref{ion_zone} must be quite 
ill-defined and ``fuzzy'' at large and medium size scales.

\subsection{Are Similar Spectral Changes Observed at Higher Stellar Latitudes?}
\label{sec:FOS4}

Our line-of-sight to  \ec\ corresponds to stellar latitudes of about 45--50\degree\ \citep{2001AJ....121.1569D,2003ApJ...586..432S} and as discussed above, spectra from this direct view show dramatic spectral changes over the past decade.  The Homunculus nebula reflects light from the central source  and allows us to view the star and its spectrum from different directions. The known geometry of the Homunculus makes it possible to directly  relate locations in the nebula to stellar latitudes 
   \citep{2003ApJ...586..432S,2001AJ....121.1569D,1999A&A...344..211Z}.
Spectra at FOS4, located near the center of the SE lobe, correspond to a stellar latitude of about 75\degree\ permitting us to observe the star's spectrum from near its polar region. (Observed delay times and 
    Doppler shifts confirm the assumed geometry, see
    \citealt{2011ApJ...740...80M}.) 
Spectra were obtained at FOS4  with {\it VLT\/} UVES from 2002--2009, with {\it Gemini\/} GMOS from 2007--2009, and with  {\it Ir\'{e}n\'{e}e du Pont\/} B\&C in 2011.

\begin{figure}
\plotone{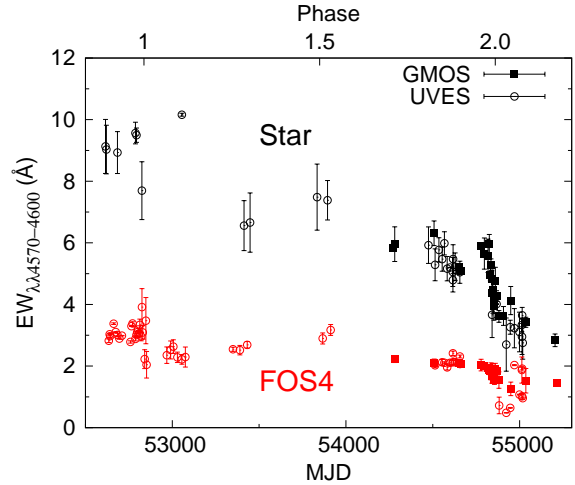}
\caption{Equivalent width of the broad \ion{Fe}{2}/\ion{Cr}{2} blend at $\lambda\lambda$4570--4600 \AA\ on the star (black symbols) and at FOS4 (red symbols) with {\it Gemini\/} GMOS and {\it VLT\/} UVES in 2002--2009. GMOS values were divided by 1.9 and UVES values by 1.7 to account for the wider spatial sampling, see text. The emission in our  direct view of the star decreased by   a factor of $\sim 3$, at FOS4 by only a factor of about 1.5--2.  \label{feII_fos4}}
\end{figure}

Figure \ref{feII_fos4} shows the equivalent width of the \ion{Fe}{2}/\ion{Cr}{2} blend at $\lambda\lambda$4570--4600 \AA\ on the star and at FOS4 with GMOS and UVES.
In 2002--2003, the equivalent width of the emission in our direct view is a factor of $\sim 3$ larger than at FOS4. It was already noted by \citet{1992A&A...262..153H} that the equivalent widths of emission lines are smaller throughout the lobes.  This fact has not been    
    fully explained, but one possible cause involves our unusual 
    line-of-sight to the star.  Our direct view of the star has more 
    extinction than the Weigelt knots located only 0\farcs3 
    away \citep{1995AJ....109.1784D,2001ApJ...553..837H}.  
    Suppose the extra obscuration occurs, for example, in a small 
    intervening dusty cloud close to the star.  Any extra emission 
    formed between us and the cloud would have a magnified effect 
    on the star's apparent spectrum, because such emission would have 
    less extinction.  In that case, the star would appear to have 
    relatively stronger emission lines than it really does.  But 
    this explanation has some obvious difficulties, and the problem 
    is too complicated to explore here.  See \citet{2003ApJ...586..432S} 
    for other related comments. \citet{2005A&A...435..303S} and \citet{2005AJ....129.1694W} also noticed the difference but without discussion.
The equivalent width in our direct view of the star declined by a factor of about 3 since 2002, while at FOS4 the decline was only by a factor of 1.5--2. After the 2009 event the strength of the emission feature was comparable at both locations.

\begin{figure}
\plotone{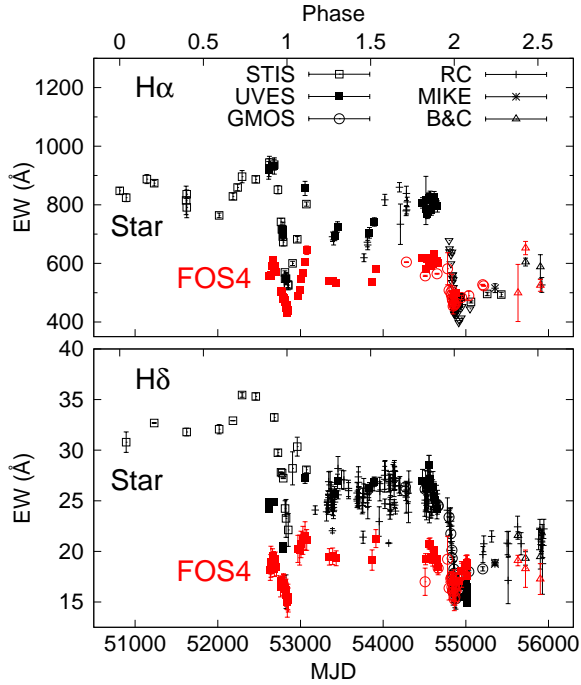}
\caption{H$\alpha$ and H$\delta$ equivalent widths in spectra of the star in direct view (black symbols) and at FOS4 (red symbols) in spectra obtained in 1998--2012. (The open black triangles turned downwards are from {\it 1.5 m CTIO} RC and Echelle observations and are retrieved from \citealt{2010AJ....139.1534R}.) The emission strengths decreased on the star by a factor of $\sim 1.5$ but not at FOS4. \label{halpha_fos4}}
\end{figure}

Similar behavior is observed in the hydrogen emission lines. Figure \ref{halpha_fos4} compares the H$\alpha$ and H$\delta$ equivalent widths in spectra of the star in direct view and reflected at FOS4 obtained with different instruments. The emission strength in spectra of the star decreased by a factor of $\sim 1.5$ since 1998 (see Section \ref{sec:long-term}), but spectra at FOS4 showed no secular changes. After the 2009 event the emission strengths were about equal at both locations. Conceivably this is a hint that the wind has become more spherical.

\begin{figure}
\plotone{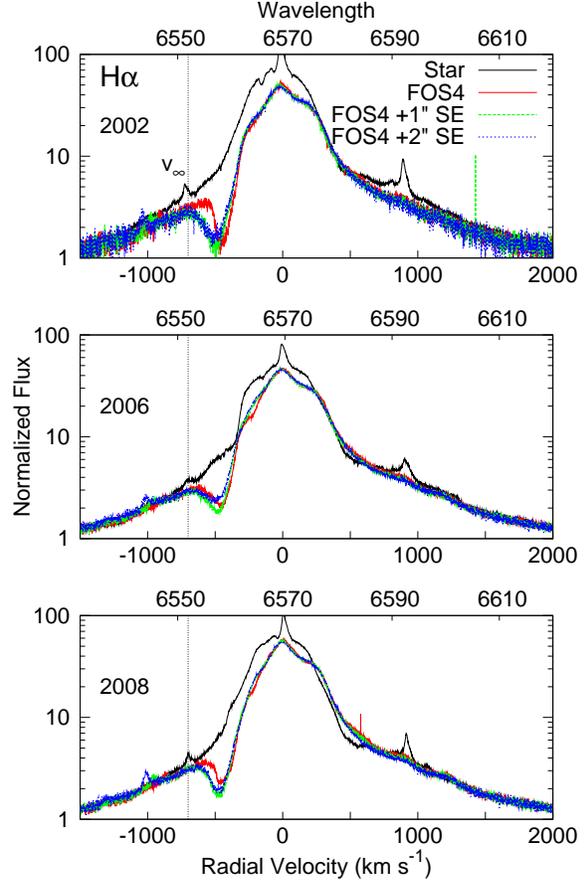}
\caption{H$\alpha$ in {\it VLT\/} UVES spectra on the star and in the SE lobe. FOS4 +1{\arcsec} and FOS4 +2{\arcsec} are extraction along the slit 1{\arcsec} and 2{\arcsec} south of FOS4. Spectra were shifted by $-100$ km s$^{-1}$ (FOS4), $-150$ km s$^{-1}$ (FOS4 +1{\arcsec}) and $-200$ km s$^{-1}$ (FOS4 +2{\arcsec}) to account for the expanding nebula. Observations in 2002 and 2008 were obtained at similar phases (0.891 and 1.886). Tracings from 2006 show the mid-cyle profile.  The maximum terminal velocity is $v_{\infty} \sim -700$ km s$^{-1}$ and the absorption feature may have weakened since 2002. \label{UVES_fos4}}
\end{figure}

\citet{2003ApJ...586..432S} reported faster terminal velocities of Balmer P Cyg absorption lines at the poles than at lower latitudes in 2000 March STIS data during \ec's normal state, which lead them to conclude that \ec's wind is faster at the poles.  They found terminal velocities of H$\alpha$ P Cyg absorption of  $v_{\infty} = -540$ km s$^{-1}$ in our direct line-of-sight view and up to $v_{\infty} = -1150$ km s$^{-1}$ in the reflected polar-on spectra. In pre-2009 event ground-based data we did not find such high velocities at the poles. Observations with GMOS starting in 2007 show terminal velocities of the H$\delta$ absorption on the order of $v_{\infty} \sim -550$ km s$^{-1}$ at all latitudes \citep{2011ApJ...740...80M}. UVES observations $\sim 200$ days before the 2003.5 and 2009 events and during mid-cycle state in 2006 show that the maximum terminal velocities for H$\alpha$ increase somewhat with higher latitude and range from  $v_{\infty} \sim -550$ to  $v_{\infty} \sim -700$ km s$^{-1}$, see Figure \ref{UVES_fos4}. The telescope acquisition of the FOS4 location has an uncertainty of $\sim \pm 0.5{\arcsec}$ and this is the likely reason that the velocity dependence observed in the 2002 and 2008 spectra is not seen in the 2006 spectra shown in the Figure. 

\begin{figure}
\plotone{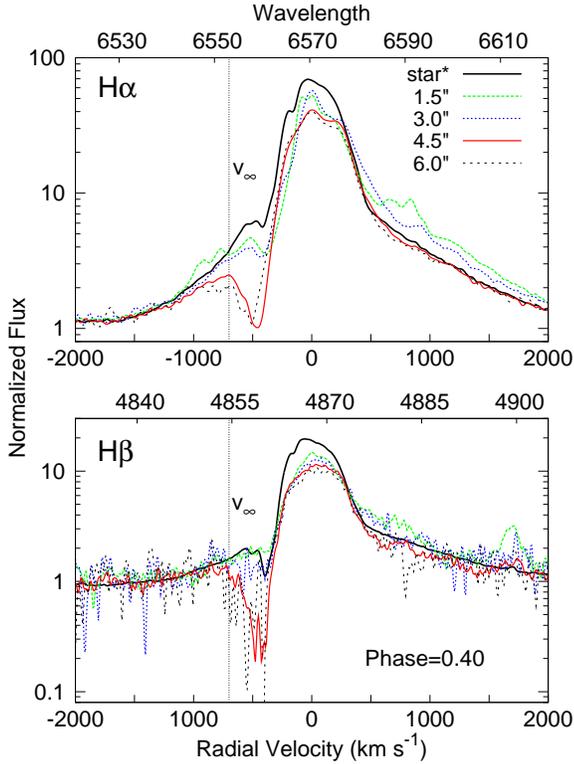}
\caption{H$\alpha$ and H$\beta$ in 2000 March {\it HST\/} STIS observations in tracings along the SE lobe (the distance in arcsec from the central source is indicated). We corrected for the different redshifts using velocities of $-12$ km s$^{-1}$ for offset position 1\farcs5, $-43$ km s$^{-1}$ for offset position 3\farcs0, $-99$ km s$^{-1}$ for offset position 4\farcs5, and $-185$ km s$^{-1}$ for offset position 6\farcs0.  The flux was normalized between $\lambda\lambda$6630--6650 \AA\ and  $\lambda\lambda$4980--5000\AA, respectively.  The terminal velocity is latitude-dependent, with the polar-on spectra showing the largest terminal velocities of $v_{\infty} \sim -700$ km s$^{-1}$. \label{STIS_fos4}}
\end{figure}

Because we did not observe terminal velocities above  $v_{\infty} = -700$ km s$^{-1}$ in our ground-based data, we reinvestigated the 2000 March STIS data used by \citet{2003ApJ...586..432S} using a different approach in aligning the spectra from several distinct locations in the Homunculus nebula.
\citet{2003ApJ...586..432S} corrected for the different redshifts throughout the SE lobe, which are due to reflection by the expanding dust, by aligning the blue side of the H$\alpha$ emission line profile at 10 times the continuum flux. In  \citet{2011ApJ...740...80M} we used, instead, several forbidden lines that are known to originate in the Weigelt knots with constant velocities much smaller than the discrepancy in question to align GMOS spectra. We cannot use the same procedure for the STIS spectra because the narrow lines cannot be as readily observed throughout the SE lobe due to the small spectral range of each exposure and the low S/N in extractions in the lobe. We therefore applied the velocities found for different locations in the SE lobe using GMOS data to the STIS spectra. The result is shown in Figure \ref{STIS_fos4}.
Using our aligning method we found maximum terminal velocities of $v_{\infty} \sim -700$ km s$^{-1}$ for H$\alpha$ and H$\beta$. 
  Admittedly $v_\infty$ is difficult to define 
     precisely in a case like this.  The lower two H$\alpha$ profiles 
     in Figure \ref{STIS_fos4} appear to show a deficit of flux between $-700$ 
     and $-950$ km s$^{-1}$, but this is not a smooth continuation
     of the main P Cyg profile.  Instead, these two examples are better 
     described as having a possible weak second component of outflow 
     with $v_\infty \sim -900$ km s$^{-1}$ rather than $-1150$ km s$^{-1}$.    
     The H$\beta$ data are noisier, but this line produces deeper 
     absorption than H$\alpha$;  and it too shows no evidence for  
     $v_\infty < -700$ km s$^{-1}$.   Figure \ref{STIS_fos4} shows a clear latitude 
     dependence, but the velocity range is less dramatic than that 
     found by \citet{2003ApJ...586..432S}. 
Unfortunately, no UV observations of the reflected polar-on spectrum exist and we therefore cannot investigate the terminal velocities of UV resonance lines at higher latitudes.

Our last observations taken in 2010 January with GMOS and in 2011 with {\it Ir\'{e}n\'{e}e du Pont\/} B\&C indicate that the absorption at the poles had weakened considerably after the 2009 event. However, since the {\it Ir\'{e}n\'{e}e du Pont\/} observations are of lower quality this has to be confirmed in future observations.

The simplest explanation for the weakening of broad stellar wind-emission features is a decrease in \ec's mass-loss rate \citep{2010ApJ...717L..22M}. The  broad stellar wind-emission features appear to be similar from all directions after the 2009 event suggesting that \ec's asymmetric wind \citep{2003ApJ...586..432S}  may have become more spherical over the last 10 years.
If the interpretation of a decrease in mass-loss rate is correct, then the effect is latitude dependent with the mass-loss rate  decreasing  less or more slowly at the higher stellar latitudes. However, \ec's wind is normally assumed to be denser at the poles \citep{2003ApJ...586..432S} and a larger decrease of the mass-loss rate at the equator would not lead to a more symmetric wind.

\subsection{Are Spectral Changes Observed at the Weigelt Knots?}
\label{sec:knots}

Spectra of the Weigelt knots show reflected light from \ec\ and narrow high-excitation emission lines \citep{1995AJ....109.1784D} now  
attributed to photoionization by a hot companion star.
Given the rapid spectral changes discussed above, and the accelerated brightening of the central star for the last 15 years \citep{2004AJ....127.2352M,2006AJ....132.2717M,2009IAUC.9094....1D}, we expect to observe spectral changes also in the nearby ejecta. For instance, an early recovery of the high-excitation emission after the 2009 event and a larger continuum flux at the Weigelt knots seem reasonable. Unfortunately, the Weigelt knots cannot be spatially resolved in ground-based observations and their observational coverage with STIS is sparse; in 2003 the pre-event phase was covered, while the recovery phase was observed during the 1998 and 2009 events. Mid-cycle observations are even rarer.

\begin{figure}
\centering
\plotone{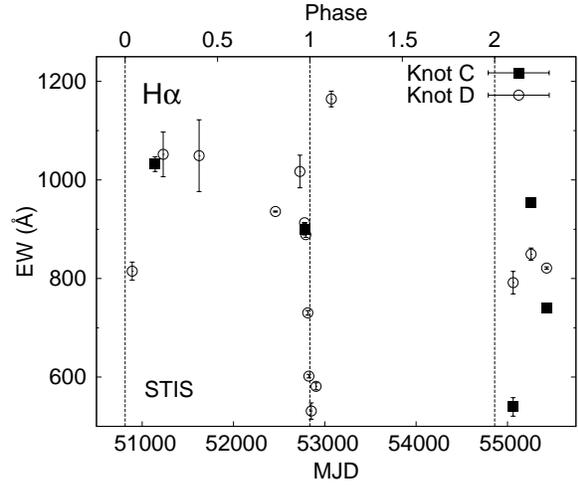}
\caption{Equivalent width of H$\alpha$ at the Weigelt knots C (filled squares) and D (open circles) over the last 2 cycles in {\it HST\/} STIS data. The equivalent width may have declined by about 10--20\% over the last decade.\label{seachange:fig:fig100}}
\end{figure}

Figure \ref{seachange:fig:fig100} shows measurements of the H$\alpha$ equivalent width at Weigelt knots C and D in STIS data for the last two cycles.\footnote{Note that the meaning of ``equivalent width''
      is unclear for the Weigelt knots.  This is because the source of continuum is
      ill-defined, mainly reflected star light but continuum emission
      in the knots may be present.}
Further observations are required to confirm the apparent long-term decrease in the emission strength of about 10--20\%. Factors such as slightly varying slit position angles, pointing, and the fact that the knots are slowly moving outwards (on the order of 0\farcs023--0\farcs044 within 10 years, see \citealt{2004ApJ...605..405S,2004AJ....127.1052D}) might play a role.  We are not concerned here with the line behavior during the events, when the emission strength drops very rapidly for a few months.

\begin{figure}
\centering
\plotone{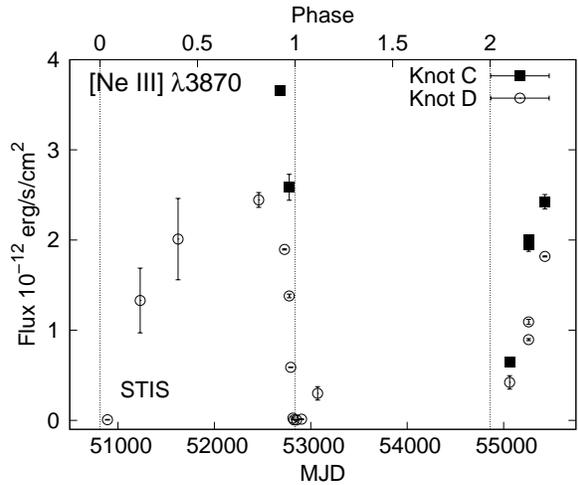}
\caption{Flux of the narrow [\ion{Ne}{3}] $\lambda$3870 emission line at Weigelt knots C (filled squares) and D (open circles) since 1998 in {\it HST\/} STIS data. The line strength may have recovered earlier after the 2009 event, see text. \label{weigeltknots:fig:fig3}}
\end{figure}

Figure \ref{weigeltknots:fig:fig3} shows the flux of the narrow [\ion{Ne}{3}] $\lambda$3870 emission on Weigelt knots C and D since 1998  (compare \citealt{2010ApJ...710..729M}).
High-excitation emission lines disappear for several months during the events, probably caused by the suppression of UV radiation from the secondary star close to periastron passage. 
Some authors have suggested that the disappearance of the high-excitation lines are caused by eclipses of a hot secondary star by the 
primary wind or wind-wind collision shock cone \citep{1997NewA....2..107D,1999ASPC..179..266I,1999ASPC..179..295S,2002A&A...383..636P,2008MNRAS.386.2330D} or due to a thermal/rotational recovery cycle \citep{1984A&A...137...79Z,2000ApJ...530L.107D,2003ApJ...586..432S,2005ASPC..332..101D}. 
Many authors now agree that a collapse of the wind-wind 
     collision structure \citep{2002ASPC..262..267D,2003ApJ...597..513S, 
2006ApJ...640..474M,2006ApJ...652.1563S,2007ApJ...661..482S,2008MNRAS.386.2330D}, and/or disturbances in 
     the primary wind \citep{1997NewA....2..387D,1999ASPC..179..304D,2003ApJ...586..432S,2006ApJ...640..474M}, are primary causes for the 
     observed spectral changes during the events.  These phenomena 
     can be triggered by the periastron passage of a companion star.

The [\ion{Ne}{3}] $\lambda$3870 emission appears to have recovered faster after the 2009 than after the 1998 event.
If \ec's wind has been decreasing in recent years, an early reappearance of the high-excitation emission lines would be expected since a lower mass-loss rate of the primary star would result in an earlier recovery of the secondary star's UV radiation output in any proposed model. However, given the poor temporal coverage of the Weigelt knots this result is not conclusive.

\begin{figure}
\centering
\plotone{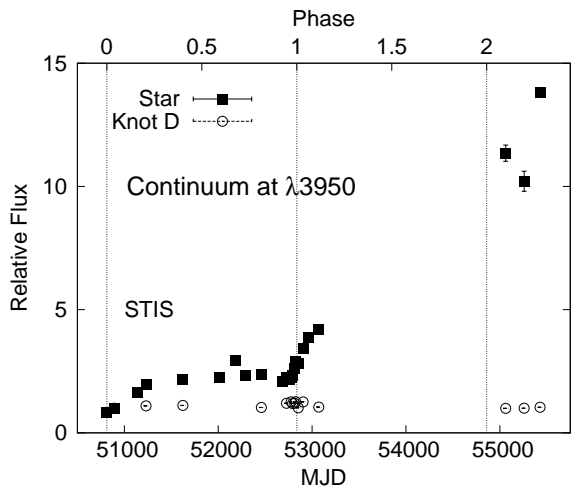}
\caption{Normalized continuum flux at $\lambda$3950 \AA\ on the star (filled squares) and at Weigelt knot D (open circles). The flux was normalized to unity on 1998 March 19 for both locations. During \ec's last two spectroscopic cycles the continuum at $\lambda \sim 4000$ \AA\ remained approximately constant at knot D, while the flux in our direct line of view rose by a factor of more than 10 between 1998 and 2010. \label{weigeltknots:cont}}
\end{figure}

Surprisingly, the continuum flux at $\sim \lambda$4000 \AA\ at Weigelt knot D is very constant for the last 10 years, see Figure \ref{weigeltknots:cont}. The Figure compares the continuum flux at the star and at Weigelt knot D. Since the stellar continuum is much brighter than the continuum at knot D, we normalized the measurements to unity on 1998 March. In 1998, the stellar continuum  at $\sim \lambda$4000 \AA\ was $\sim 5$ times as bright as on the nearby knot D. The central source then brightened tremendously (see also Figure 1 in \citealt{2011ApJ...740...80M} for {\it HST\/}  UV photometry). In 2010 August, the stellar continuum was about 60--70 times brighter than the continuum at knot D, which remained practically constant. This is quite unexpected.
However, the rapid brightening of the central star is largely caused by a decrease in the circumstellar extinction; the innermost dust is being destroyed or the dust-formation rate has slowed. Our direct view of the star appears to have more circumstellar extinction than the average line-of-sight \citep{1995AJ....109.1784D} and the brightening of the central star may not be equal in all directions.

\section{IMPLICATIONS FOR ETA CAR'S MASS-LOSS RATE}
\label{sec:massloss}

The spectral changes described in this paper suggest that \ec's wind density decreased and that the ionization structure of the inner wind changed. The changes appear to be dependent on the stellar latitude. Eta Car's wind may be more spherical now than 10 years ago. However, the nature of these spectral changes cannot be easily explained.

In \ec's ``normal'' state, Balmer P Cyg absorption is strong at the poles and weak or absent along our line 
  of sight, near stellar latitude $\sim$ 45\degree.
  It is therefore thought that \ec's wind density is higher 
   at the poles \citep{2003ApJ...586..432S}, where it may resemble 
   the spherical model described by \citet{2001ApJ...553..837H}.    
   At lower latitudes, in this view, the wind is less dense, which 
   implies stronger ionization and much weaker Balmer absorption.   
   (The column density $N(H^{0},n=2)$ is small there because 
   $N(H^+ + H^0)$ and the ratio $N(H^0,n=2)/N(H^+)$ are both smaller 
   than they are at the poles.) A complex photoionization structure of the primary wind regulated by the secondary star \citep{2010AJ....139.1534R} or a wind cavity model \citep{2011arXiv1111.2280M} may provide additional or alternative explanations. 
   
During the events, Balmer P Cyg absorption also appears at lower latitudes and
the rapidly changing profiles indicate changes in \ec's wind ionization structure on very short time scales of only days. \citet{2003ApJ...586..432S} proposed that a minor mass ejection leads to a temporary increase in \ec's wind density in the equatorial regions resulting in hydrogen recombination.
However, \ec's wind might be close to a regime where a small change in its wind parameters may lead to transitions between fully ionized and recombined hydrogen in the wind. This may be the case during the events, when the radiation of the secondary might cause a rapid transition between these two states and the ionization structure of \ec's wind might temporarily change \citep{2010AJ....139.1534R}. \citet{2011arXiv1111.2280M} found that a wind cavity in the dense primary wind caused by the secondary star may provide an explanation for the deepening of H$\alpha$ absorption in our line-of-sight during the events. Observations favoring the latter explanations are the constant terminal wind velocities in UV resonance lines during the 2003.5 event (see Figure \ref{UV_Si1527}) and the appearance of \ion{He}{1} absorption at higher stellar latitudes for a few months before the 2009 event \citep{2011ApJ...740...80M}. UVES spectra before the 2003.5 event, starting at phase 0.9, show also strong \ion{He}{1} absorption at the pole. This occurrence is not accounted for by a shell-ejection model.

The long-term weakening of \ion{H}{1} emission in \ec's wind may be explained with a decrease in mass-loss rate, while the constant \ion{He}{1} emission strength is probably due to competing effects of changes in the helium ionization, which is due mainly to UV from the hot companion star. Long-term changes in the \ion{H}{1} and \ion{He}{1} P Cyg absorption lines are related to changes in the ionization structure of \ec's wind and likely caused by alterations in the mass-loss rate. For example,  \citet{1997A&A...326.1117N} demonstrated that the variability of the \ion{H}{1} and \ion{He}{1} line profiles in P Cygni resulted from changes in the ionization of its wind. 

Let us assume that the observed weakening of broad stellar wind features is primarily caused by a decreasing mass-loss rate, which seems natural for \ec's long-term recovery.  A decrease in mass-loss rate is consistent with the accelerated secular brightening trend in {\it HST\/} images and spectroscopy \citep{1999AJ....118.1777D,2004AJ....127.2352M,2006AJ....132.2717M}
as well as other recent observational evidence \citep{2005AJ....129..900D, 2006AJ....132.2717M,
2008AJ....135.1249H, 2009ApJ...701L..59K,2010AJ....139.2056M,2010ApJ...725.1528C}.   

Previous mass-loss rate estimates for \ec\ range from $3 \times 10^{-4}$ to $10^{-3} M_\odot$ yr$^{-1}$.  
\cite{1995AJ....109.1784D} estimated the mass-loss rate based on the H$\beta$ emission line and found $6 \times 10^{-4}$ to $3 \times 10^{-3} M_\odot$ yr$^{-1}$, with a most likely value of $1 \times 10^{-3} M_\odot$ yr$^{-1}$.
\citet{2001ApJ...553..837H} also found a mass-loss rate of $\sim 10^{-3} M_\odot$ yr$^{-1}$ by fitting the optical emission spectrum with a non-LTE line blanketed code. Radio observations at 8 and 9 GHz  indicate mass-loss rates of $3 \times 10^{-4} M_\odot$ yr$^{-1}$ \citep{1994ApJ...429..380W} and millimeter observations resulted in $2.4 \times 10^{-3} M_\odot$ yr$^{-1}$ \citep{1995A&A...297..168C}.
All those estimates are based on simplified, spherical models and are only order of magnitude estimates.\footnote{The 8--9 GHz observations see inhomogeneous material far outside the normal stellar wind because the opaque region at those frequencies probably includes all of the Weigelt knots. }  
Mass-loss rates obtained from optical observations are higher than from X-ray models, which find mass-loss rates of about $3 \times 10^{-4} M_\odot$ yr$^{-1}$ \citep{1999ApJ...524..983I,2001ApJ...547.1034C,2002A&A...383..636P}.  This discrepancy might be reduced if clumping is taken into account since the mass-loss rates determined from $\rho^2$ diagnostics may have been systematically overestimated by up to an order of magnitude \citep{2006ApJ...637.1025F}.

In this paper we did not attempt to estimate the absolute mass-loss rate of \ec\ because there are too many unknowns such as the latitudinal dependence and clumping of the wind. Instead, we adopted the method by \citet{1988ApJ...326..356L} which relates the H$\alpha$ luminosity to stellar mass-loss rate, stellar radius, velocity law, and effective temperature, to roughly estimate the change in mass-loss rate over the last 10 years.  Assuming that only the mass-loss rate is responsible for the observed changes in H$\alpha$ flux, we find that {\it the mass-loss rate declined by a factor of 2--3 between 1999 and 2010.} Note: we do find absolute mass-loss rates on the right order of magnitude, i.e. $10^{-4}$--$10^{-3} ~M_\odot$ yr$^{-1}$. A full theoretical analysis requires expert codes and new models updating \cite{2001ApJ...553..837H} are needed.

A decrease in the mass-loss rate by a factor of 2--3 is consistent  with estimates based on the X-ray light curve. The early exit from \ec's 2009 X-ray minimum suggests a decrease in mass-loss rate by a factor of 2 compared to previous events \citep{2009ApJ...701L..59K}. A decrease in mass loss rate by a factor of 2 also results from the decline in 2--10 keV X-ray flux by $\sim 30$\% between 2000 and 2011.\footnote{See http://asd.gsfc.nasa.gov/Michael.Corcoran/ eta\_car/etacar\_rxte\_lightcurve/index.html for the 2--10 keV X-ray lightcurve obtained with the RXTE/PCA PCU2 Layer 1. The X-ray flux of the
colliding winds is proportional to $\dot{M}_{\eta Car}^{1/2}$.}   \citet{2010ApJ...725.1528C} estimated a factor of 4 decrease in the mass loss rate between 2000 and 2006, which may 
be too excessive as the comparison was made based on the fluxes obtained nearly at a local maximum 
in 2000 and a local minimum in 2006. Corcoran et al.\ also suggested changes 
in the plasma temperature of the colliding wind shocks, which makes it difficult to assess what 
physical quantities -- other than  mass loss rate of \ec\ -- may have changed. 

A decreasing mass-loss rate could also potentially explain the deepening of \ion{He}{1} and \ion{N}{2} absorption over the last decade. Eta Car's wind may be in a stage where even a modest change in mass-loss rate can have a large impact on the wind ionization structure and a decrease in mass-loss rate may cause helium to become ionized in a larger fraction of the wind at low latitudes.

A dramatic drop in \ec's mass-loss rate mainly at the equatorial regions, however, leads to a significant conflict. Theories of equatorial gravity darkening in massive rotating stars \citep{2000A&A...361..159M,2001A&A...372L...9M,2005ASPC..332..169O} result in asymmetric winds with stellar wind densities and terminal wind velocities being larger at the poles, and the generally accepted hypothesis is that \ec's mass-loss rate was higher at the poles than at the equator \citep{2003ApJ...586..432S}.  However, recent data imply a more spherical wind; the terminal velocities of Balmer P Cyg absorption appear to be fairly constant at all latitudes and emission strengths are equal from all directions. This cannot easily be explained alongside with a rapid decrease in mass-loss rate mainly at the equator.
However, given the observational evidence presented above, the interpretation of a latitude-dependent wind caused by rapid stellar rotation might not be correct. 
Alternatives to the decreasing-wind interpretation include, e.g., a change in the latitude-dependence of the wind, changes in the velocity field shape, or the model favored
by \citet{2009ApJ...701L..59K}, who propose that a small change in wind properties could be amplified by tidal interactions.
More detailed analysis and future observations in the next years are necessary. We can only state here, that \ec's wind has changed considerably over the last decade but any explanation of the nature of these changes is not straightforward.
   
      As noted in earlier papers, \ec\ may now be returning
      to a state like that observed three centuries ago, with
      a nearly transparent wind   \citep{2006AJ....132.2717M,  
      2010ApJ...717L..22M}.  Conceivably, however, it may
      already have reached that state.     In 1998 its opaque wind
      had a pseudo-photospheric temperature of 9000--14,000 K  
      \citep{2001ApJ...553..837H}.   Figure 1 in  
      \citet{1987ApJ...317..760D} indicates that a factor of 2 or 3
      decrease in the wind density should probably have raised
      the apparent temperature to 20,000 K or more.  
      (Modernized opacities do not alter this relative
      statement.)  According to an argument based on
      the star's bolometric magnitude compared to the visual
      magnitude seen by Halley in 1677, the color temperature long before the Great Eruption 
     was most likely about 20,000 to 25,000 K
      \citep{2012eta_Kris}.  This
      value may represent either the star's true effective
      temperature, or else a marginally opaque wind.  If this
      reasoning is valid, perhaps the circumstellar extinction
      is the only remaining difference between the star's
      appearance today and that seen 150 years before the
      Great Eruption.  One implication is that the near-future
      development cannot safely be predicted merely by extrapolating
      from the past decade.

\section{SUMMARY}
\label{sec:discussion}

In this paper we analyzed spectral data obtained with several instruments between 1998 and 2012.
We confirmed the spectral changes in the wind emission lines first reported  by \citet{2010ApJ...717L..22M};  {\it HST} STIS spectra obtained in 2010 August, $\sim 170$ days after the first discovery, are comparable to the observations in 2010 March.
Furthermore, we analyzed the long-term development of spectral changes in our direct line-of-sight view of  the star, at FOS4, and the Weigelt knots.

Eta Car's recent spectral changes involve both emission and absorption lines:
\begin{enumerate}
\item Broad stellar wind-emission features in our line-of-sight to the star have decreased by factors of 1.5--3 relative to the continuum within the last 10 years. These changes occurred gradually and are dependent on the viewing angle; spectra at higher stellar latitudes and from the outlying ejecta show smaller changes.
The simplest explanation  is a decrease in $\eta$ Car's
primary wind density.
However, the decrease in wind density appears to be latitude dependent, with emission features showing much less change at higher latitudes. After the 2009 event, emission line strengths are now very similar in our direct line-of-sight view and in the reflected polar-on spectrum at FOS4 suggesting a more spherical wind and/or a more uniform distribution of circumstellar extinction.
\item High-excitation \ion{He}{1} and \ion{N}{2} absorption lines strengthened gradually over the last decade indicating a change in \ec's wind ionization structure. Hydrogen P Cyg absorption at FOS4 might have weakened after the 2009 event.  The terminal velocity of hydrogen P Cyg lines was found to be similar at all stellar latitudes. Those findings provide additional clues for a more spherical wind. 
\end{enumerate}

The observational results presented here are difficult to reconcile with a decrease in mass-loss rate primarily at lower stellar latitudes since it is generally assumed that \ec's wind had higher densities at the poles \citep{2003ApJ...586..432S}. 
Our observations may be more readily reconciled with alternative explanations for latitude-dependent spectral features, such as a complex ionization structure of  \ec's wind modulated by the secondary star's UV radiation \citep{2010AJ....139.1534R} or the presence of a wind cavity in the primary wind caused by the secondary star \citep{2010ApJ...716L.223G,2011arXiv1111.2280M}.

Using H$\alpha$ emission and the method by \citet{1988ApJ...326..356L} we found that \ec's mass-loss rate decreased by a factor of 2--3 between 1999 and 2010. A decrease in mass-loss rate on the order of 2--3 is consistent with changes in the  X-ray light curve
\citep{2009ApJ...701L..59K,2010ApJ...725.1528C}. We did not attempt to derive the absolute value with any accuracy because there are too many unknown factors, such as latitudinal dependence and clumping of the wind. New theoretical models updating \cite{2001ApJ...553..837H} are needed.  

Observations in 2012 and 2013 will be extremely valuable to further analyze the nature of the spectral changes in \ec's wind.
It is of great importance to monitor the star consistently since spectral changes may occur on time scales of only weeks to months. For the long-term recovery of \ec\ it is important to investigate if the wind will further decline or if it will stabilize or even recover to its former strength. But by mid-2013, the onset of the  next event will dominate the spectrum, so observations in 2012 are needed. The last three events all differed from each other and considering the long-term spectral changes described in this paper we can expect many interesting new results from \ec's 2014.5 event.

{\it Acknowledgement}
We thank the staff and observers of the Gemini-South Observatory in La Serena for their help in preparing and conducting the observations, and Beth Perriello at STScI for assistance with {\it HST\/} observing plans. We also thank Otmar Stahl and Kerstin Weis for their effort in planning and obtaining the UVES spectra. 
AM was co-funded under the Marie Curie Actions of the European Commission (FP7-COFUND).
MTR received partial support from Center for Astrophysics FONDAP and PB06 CATA (CONICYT).

\epsscale{1}

\newpage

\begin{deluxetable}{lcccccc}
\tabletypesize{\scriptsize}
\tablecaption{Equivalent Widths of Broad Stellar Wind-emission Features\tablenotemark{a} (1998--2012)\label{tab:table1}}
\tablewidth{0pt}
\tablehead{
\colhead{Name\tablenotemark{b}} &
\colhead{Date} &
\colhead{MJD}   &
\colhead{Phase} &
\colhead{EW$_{\lambda\lambda4570-4600\tablenotemark{c}}^{Star}$} &
\colhead{EW$_{\lambda\lambda4614-4648\tablenotemark{c}}^{Star}$} &
\colhead{EW$_{\lambda\lambda4570-4600\tablenotemark{c}}^{FOS4}$}      \\
\colhead{} &
\colhead{(UT)} &
\colhead{}   &
\colhead{} &
\colhead{(\AA)} &
\colhead{(\AA)} &
\colhead{(\AA)}  
}
   \startdata
\multicolumn{7}{c}{{\it HST\/} STIS}\\
c821 	&	1998 Mar 19 	&	50891.4	&	0.038	&	11.02 $\pm$ 	0.05 &	8.84 $\pm$	0.01 &	\nodata \\
c914 	&	1999 Feb 21	&	51230.5	&	0.206	&	16.51 $\pm$0.06 	&	12.37 $\pm$ 0.10	& \nodata	\\
cA22 	&	2000 Mar 20 	&	51623.8	&	0.400	&	15.99 $\pm$ 0.42	&	11.55 $\pm$ 0.21	&	\nodata	\\
cB29 	&	2001 Apr 17	&	52016.8	&	0.595	&	11.81 $\pm$ 0.13	&	8.90 $\pm$ 0.21	&	\nodata \\
cC05 	&	2002 Jan 20	&	52294.0	&	0.732	&	14.40 $\pm$ 0.06	&	10.73 $\pm$ 0.04	& \nodata	\\
cC51 	&	2002 Jul 04	&	52459.5	&	0.813	&	13.72 $\pm$ 0.46	&	10.46 $\pm$ 0.28	&	\nodata \\
cD12 	&	2003 Feb 13	&	52683.1	&	0.924	&	9.39 $\pm$ 0.01	&	7.72 $\pm$ 0.10	&	\nodata \\
cD24 	&	2003 Mar 29	&	52727.3	&	0.946	&	9.31 $\pm$0.03	&	7.49 $\pm$ 0.11	& \nodata	\\
cD34 	&	2003 May 05	&	52764.3	&	0.964	&	10.10 $\pm$ 0.07	&	7.98 $\pm$ 0.25	& \nodata	\\
cD37 	&	2003 May 19	&	52778.5	&	0.971	&	10.96 $\pm$ 0.06	&	8.58 $\pm$ 0.04	& \nodata	\\
cD41 	&	2003 Jun 01	&	52791.7	&	0.978	&	12.47 $\pm$ 0.29	&	9.69 $\pm$ 0.01	&	\nodata \\
cD47 	&	2003 Jun 23	&	52813.8	&	0.989	&	11.78 $\pm$ 0.11	&	10.38 $\pm$ 0.23	&	\nodata \\
cD51 	&	2003 Jul 05	&	52825.4	&	0.994	&	10.89 $\pm$ 0.81	&	8.87 $\pm$ 0.25	&	\nodata \\
cD58 	&	2003 Aug 01	&	52852.4	&	1.008	&	12.50 $\pm$ 0.44	&	9.90 $\pm$ 0.39	&	\nodata \\
cD72 	&	2003 Sep 22	&	52904.3	&	1.033	&	10.31 $\pm$ 0.10	&	8.36 $\pm$ 0.03	&	\nodata \\
cD88	&	2003 Nov 17	&	52960.6	&	1.061	&	11.95 $\pm$ 0.14	&	9.25 $\pm$ 0.21	&	\nodata \\
cE18	&	2004 Mar 07 &	53071.2	&	1.116	&	9.20 $\pm$ 0.27	&	7.69 $\pm$ 0.12	&	\nodata \\
cJ49	&	2009 Jun 30	&	55012.1	&	2.075	&	3.42 $\pm$ 0.27	&	3.39 $\pm$ 0.34 &	\nodata	\\
cJ63	&	2009 Aug 19	&	55062.0	&	2.100	&	2.58 $\pm$ 0.21 	&	3.82 $\pm$ 0.03	&	\nodata \\
cJ93 & 2009 Dec 06 & 55171.6 &2.154 & 4.3 $\pm$ 0.15  & 3.78 $\pm$ 0.32 & \nodata \\
cK16 	&	2010 Mar 03	&	55258.6	&	2.197	&	5.07 $\pm$ 0.10 &	4.19 $\pm$ 0.12	&	\nodata \\
cK63 	&	2010 Aug 20	&	55428.3	&	2.281	&	4.64 $\pm$ 0.18 	&	4.18 $\pm$ 0.05	&	\nodata \\
cK81 &	2010 Oct 26 & 	55495.1 & 2.314 &	4.32 $\pm$ 0.17&	3.35 $\pm$ 0.01& \nodata\\	
\tableline  
\multicolumn{7}{c}{{\it VLT\/} UVES}\\
uC93 & 2002 Dec 7 & 52615.3 & 0.890 & 15.52$\pm$1.49 & 12.97$\pm$1.21 &  \nodata \\
uC95 & 2002 Dec 12 & 52620.3 & 0.893 & 15.35$\pm$1.34 & 12.69$\pm$1.01  & \nodata \\
uC98 & 2002 Dec 26 & 52634.4 & 0.900 & \nodata & \nodata & 4.80$\pm$0.10  \\
uD00 & 2002 Dec 31 &  52639.3 & 0.902 & \nodata & \nodata & 5.15$\pm$0.06  \\
uD00 & 2003 Jan 3 & 52642.3 & 0.904 & \nodata & \nodata & 5.05$\pm$0.09  \\
uD05 & 2003 Jan 23 & 52662.4 & 0.914 & \nodata & \nodata & 5.74$\pm$0.05  \\
uD09 & 2003 Feb 4 & 52674.4 & 0.920 & \nodata & \nodata & 5.28$\pm$0.02  \\
uD12 & 2003 Feb 14 & 52684.1 & 0.924 & 15.18$\pm$1.16 & 12.71$\pm$0.87 & 5.12$\pm$0.09 \\
uD15 & 2003 Feb 25 & 52695.3 & 0.930 & \nodata & \nodata & 4.92$\pm$0.03 \\
uD18 & 2003 Mar 12 & 52710.0 & 0.937 & \nodata & \nodata & 5.08$\pm$0.01 \\
uD33 & 2003 Apr 30 & 52759.1 & 0.962 & \nodata & \nodata & 4.73$\pm$0.07 \\
uD33 & 2003 May 5 & 52765.0 & 0.964 &  \nodata & \nodata & 5.60$\pm$0.02 \\
uD36 & 2003 May 12 & 52771.2 & 0.967 & \nodata &\nodata  & 5.75$\pm$0.10 \\
uD40 & 2003 May 29 & 52788.1 & 0.976 & 16.26$\pm$0.60 & 13.34$\pm$0.44 &  4.90$\pm$0.11 \\
uD42 & 2003 Jun 3 & 52794.0 & 0.979 & 16.14$\pm$0.39 & 12.81$\pm$0.33 & 5.11$\pm$0.11 \\
uD42 & 2003 Jun 8 & 52798.0 & 0.981 & \nodata &\nodata   & 5.26$\pm$0.06 \\
uD45 & 2003 Jun 13 & 52803.0 & 0.983 & \nodata & \nodata &  5.42$\pm$0.08 \\
uD45 & 2003 Jun 17 & 52808.0 & 0.986 & \nodata & \nodata &  5.18$\pm$0.15 \\
uD47 & 2003 Jun 22 & 52813.0 & 0.988 & \nodata & \nodata &  5.68$\pm$0.32 \\
uD49 & 2003 Jun 30 & 52821 & 0.992 & \nodata & \nodata &  5.03$\pm$0.24 \\
uD51 & 2003 Jul 5 & 52825.0 & 0.994 & 13.08$\pm$1.59 & 10.35$\pm$1.19 &  6.65$\pm$1.02 \\
uD51 & 2003 Jul 9 & 52830.0 & 0.997 & \nodata & \nodata &  5.25$\pm$0.37 \\
uD54 & 2003 Jul 21 & 52841.0 & 1.002 & \nodata & \nodata &  3.78$\pm$0.52 \\
uD57 & 2003 Jul 27 & 52848.0 & 1.005 & \nodata & \nodata &  5.91$\pm$1.28 \\
uD57 & 2003 Aug 1 & 52852.0 & 1.007 & \nodata & \nodata &  3.47$\pm$0.73 \\
uD90 & 2003 Nov 25 & 52968.3 & 1.065 & \nodata & \nodata &  4.00$\pm$0.45 \\
uD96 & 2003 Dec 17 & 52990.3 & 1.076 & \nodata & \nodata &  4.29$\pm$0.50 \\
uE00 & 2004 Jan 2 & 53006.3 & 1.084 & \nodata & \nodata &  4.47$\pm$0.38 \\
uE07 & 2004 Jan 25 & 53029.3 & 1.095 & \nodata & \nodata &  3.88$\pm$0.29 \\
uE14 & 2004 Feb 20 & 53055.1 & 1.108 & 17.27$\pm$0.08 & 16.15$\pm$0.06 & 3.77$\pm$0.27 \\
uE19 & 2004 Mar11 & 53075.1 & 1.118 & \nodata & \nodata & 3.90$\pm$0.55 \\
uE94 & 2004 Dec 10 & 53349.3 & 1.253 & \nodata & \nodata & 4.33$\pm$0.11 \\
uF05 & 2005 Jan 19 & 53389.2 & 1.273 & \nodata & \nodata & 4.28$\pm$0.26 \\
uF12 & 2005 Feb 12 & 53413.4 & 1.285 & 11.15$\pm$1.38 & 8.71$\pm$1.04 & \nodata \\
uF17 & 2005 Mar 2 & 53431.3 & 1.294 & \nodata & \nodata & 4.57$\pm$0.18 \\
uF21 & 2005 Mar 19 & 53448.1 & 1.302 & 11.32$\pm$1.64 & 8.75$\pm$1.23 & \nodata  \\
uG27 & 2006 Apr 9 & 53834.1 & 1.493 & 12.72$\pm$1.82 & 9.77$\pm$1.35 & \nodata \\
uG36 & 2006 May 11 & 53866.0 & 1.509 & \nodata & \nodata & 4.92$\pm$0.31 \\
uG43 & 2006 Jun 8 & 53894.0 & 1.523 & 12.55$\pm$1.09 & 9.06$\pm$0.80 &  \nodata \\
uG48 & 2006 Jun 26 & 53912.1 & 1.531 & \nodata & \nodata & 5.39$\pm$0.30 \\
uI02 & 2008 Jan 10 & 54475.3 & 1.810 & 10.07$\pm$1.01 & 9.36$\pm$0.79 &  \nodata \\
uI13 & 2008 Feb 17 & 54513.3 & 1.829 & 8.99$\pm$0.88 & 8.35$\pm$0.69  &  3.44$\pm$0.13 \\
uI19 & 2008 Mar 10 & 54535.3 & 1.839 & 9.81$\pm$0.67 & 9.47$\pm$0.53 & \nodata \\
uI24 & 2008 Mar 29 & 54554.3 & 1.849 & 9.30$\pm$0.67 & 8.88$\pm$0.53 &  3.61$\pm$0.16 \\
uI28 & 2008 Apr 11 & 54567.0 & 1.855 & 10.17$\pm$0.64 & 9.76$\pm$0.51 & 3.58$\pm$0.05 \\
uI32 & 2008 Apr 27 & 54583.0 & 1.863 & 8.78$\pm$0.67 & 8.24$\pm$0.53 & 3.33$\pm$0.15 \\
uI36 & 2008 May 12 & 54599.0 & 1.871 & 8.84$\pm$0.01 &8.31$\pm$0.01  & 3.58$\pm$0.16 \\
uI41 & 2008 May 28 & 54615.0 & 1.879 & 8.59$\pm$0.80 &8.06$\pm$0.64 & \nodata \\
uI41 & 2008 May 30 & 54616.0 & 1.879 & 8.14$\pm$0.65 &7.67$\pm$0.52  & 4.11$\pm$0.12 \\
uI41 & 2008 May 31 & 54617.1 & 1.880 & 9.31$\pm$0.79 &9.03$\pm$0.63  & 3.61$\pm$0.13 \\
uI44 & 2008 Jun 11 & 54629.0 & 1.886 & 8.88$\pm$0.77 & 8.19$\pm$0.61 & 3.61$\pm$0.10 \\
uI52 & 2008 Jul 9 & 54656.0 & 1.899 & 8.58$\pm$0.61 & 8.45$\pm$0.49 &  3.93$\pm$0.08 \\
uI52 & 2008 Jul 10 & 54657.1 & 1.900 & \nodata & \nodata & 3.51$\pm$0.14 \\
uJ03 & 2009 Jan 10 & 54841.4 & 1.991 & 6.23$\pm$1.26 & 7.67$\pm$1.06 & \nodata \\
uJ07 & 2009 Jan 25  & 54856.2 & 1.998 & \nodata & \nodata & 2.65$\pm$0.19 \\
uJ10 & 2009 Feb 5 & 54867.3 & 2.004 & 6.82$\pm$0.01 & 7.35$\pm$0.01 &  \nodata\\
uJ14 & 2009 Feb 20 & 54882.2 & 2.011 & \nodata & \nodata & 1.23$\pm$0.46 \\
uJ25 & 2009 Apr 2 & 54923.2 & 2.031 & 4.59$\pm$1.47 & 5.09$\pm$1.21 & 0.82$\pm$0.00 \\
uJ31 & 2009 Apr 25 & 54946.1 & 2.043 & 5.55$\pm$0.39 & 5.96$\pm$0.31  &  1.09$\pm$0.02 \\
uJ38 & 2009 May 19 & 54970.0 & 2.054 & 5.49$\pm$1.08 & 5.83$\pm$0.88  &  1.83$\pm$0.02 \\
uJ46 & 2009 Jun 17 & 54999.1 & 2.069 & 5.28$\pm$1.09 & 5.63$\pm$0.90 & 3.18$\pm$0.72 \\
uJ50 & 2009 Jun 30 & 55013.0 & 2.076 & 5.67$\pm$0.38 & 5.55$\pm$0.33 & 4.17$\pm$0.29 \\
uJ50 & 2009 Jul 1 & 55014.0 & 2.076 & 5.02$\pm$0.98 & 4.94$\pm$0.80 & 1.73$\pm$0.17 \\
uJ50 & 2009 Jul 2 & 55015.0 & 2.077 & 6.20$\pm$0.44 & 6.37$\pm$0.37 & 3.26$\pm$0.07 \\
uJ51 & 2009 Jul 5 & 55018.0 &	2.078 &	4.68$\pm$0.90 & 4.57$\pm$0.73  & 1.63$\pm$0.06\\
\tableline
\multicolumn{7}{c}{{\it Gemini\/} GMOS}  \\
gH45	&	2007 Jun 16	&	54268.0	&	1.707	&	11.10	$\pm$	0.20	&	10.23	$\pm$	0.19	&	\nodata	\\
gH49	&	2007 Jun 30	&	54281.0	&	1.714	&	11.32	$\pm$	1.07	&	9.70	$\pm$	0.74	&	4.23$\pm$0.21	\\
gI11	&	2008 Feb 11	&	54507.4	&	1.826	&	11.99	$\pm$	0.76	&	11.44	$\pm$	0.36	&	4.02$\pm$0.22	\\
gI50	&	2008 Jul 05	&	54652.0	&	1.897	&	9.88	$\pm$	0.10	&	8.58	$\pm$	0.45	&	4.01$\pm$0.04	\\
gI54	&	2008 Jul 17	&	54665.0	&	1.904	&	9.65	$\pm$	0.04	&	8.78	$\pm$	0.03	&	3.92$\pm$0.00	\\
gI85	&	2008 Nov 08	&	54778.3	&	1.960	&	11.21	$\pm$	0.00	&	11.81	$\pm$	0.00	&	3.86$\pm$0.36	\\
gI90	&	2008 Nov 27	&	54797.3	&	1.969	&	10.74	$\pm$	0.96	&	10.47	$\pm$	1.86	&	3.83$\pm$0.10	\\
gI96	&	2008 Dec 18	&	54818.3	&	1.979	&	10.60	$\pm$	1.03	&	10.85	$\pm$	1.15	&	3.67$\pm$0.20	\\
gI98 &	2008 Dec 25	&	54825.3	&	1.983	&	11.35	$\pm$	0.57	&	11.63	$\pm$	0.95	&	3.59$\pm$0.26 \\
gI99	&	2008 Dec 31	&	54831.3	&	1.986	&	9.44	$\pm$	0.38	&	10.26	$\pm$	0.14	&	3.50$\pm$0.28	\\
gJ01	&	2009 Jan 04	&	54835.3	&	1.988	&	10.03	$\pm$	0.38	&	10.43	$\pm$	0.86	&	3.47$\pm$0.53 \\
gJ02	&	2009 Jan 09	&	54840.2	&	1.990	&	9.14	$\pm$	0.54	&	10.08	$\pm$	0.80	&	3.17$\pm$0.43 \\
gJ03	&	2009 Jan 12	&	54843.3	&	1.992	&	8.33	$\pm$	0.28	&	9.32	$\pm$	0.08	&	3.49$\pm$0.44 \\
gJ04	&	2009 Jan 15	&	54846.2	&	1.993	&	8.48	$\pm$	1.42	&	8.72	$\pm$	1.68	&	3.24$\pm$0.30	\\
gJ05 &	2009 Jan 21	&	54852.3	&	1.996	&	7.51	$\pm$	0.88	&	8.49	$\pm$	0.69	&	2.95$\pm$0.22	\\
gJ06	&	2009 Jan 24	&	54855.3	&	1.998	&	7.94	$\pm$	0.52	&	8.27	$\pm$	0.77	&	3.08$\pm$0.42	\\
gJ07	&	2009 Jan 29	&	54860.4	&	2.000	&	9.04	$\pm$	0.85	&	8.70	$\pm$	1.15	&	3.63$\pm$0.34 \\
gJ09	&	2009 Feb 05	&	54867.2	&	2.004	&	8.12	$\pm$	0.33	&	8.49	$\pm$	0.72	&	3.52$\pm$0.11	\\
gJ13	&	2009 Feb 19	&	54881.2	&	2.011	&	6.86	$\pm$	0.36	&	7.54	$\pm$	0.68	&	2.94$\pm$0.51	\\
gJ20	&	2009 Mar 17	&	54907.3	&	2.023	&	6.89	$\pm$	0.58	&	6.94	$\pm$	0.79	&	\nodata	\\
gJ32	&	2009 Apr 28	&	54949.1	&	2.044	&	7.84	$\pm$	0.87	&	7.96	$\pm$	1.39	&	2.39$\pm$0.42	\\
gJ56	&	2009 Jul 23	&	55036.0	&	2.087	&	6.50	$\pm$	0.31	&	6.18	$\pm$	0.86	&	2.89$\pm$0.76	\\
gK02 &	2010 Jan 08	&	55204.3	&	2.170	&	5.39	$\pm$	0.39	&	5.09	$\pm$	0.58	&	\nodata \\
gK05 & 2010 Jan 20 & 55216.3 & 	2.176 & \nodata & \nodata &	2.74$\pm$0.01 \\
\tableline
\multicolumn{7}{c}{{\it Magellan II\/} MIKE}\\
\nodata & 2010 Jun 4 & 55352.62 & 2.244 & 6.95$\pm$0.36 & 3.84$\pm$0.55 & \nodata \\
\tableline
\multicolumn{7}{c}{{\it Ir\'{e}n\'{e}e du Pont\/} B\&C}\\
\nodata & 2011 Feb 25 & 55629.92 & 2.381 & 6.48$\pm$0.12 & 4.94$\pm$0.08 & 3.49$\pm$0.48 \\						
\nodata & 2011 Jun 8 & 55720.95 & 2.426 & 4.42$\pm$0.417 & 5.05$\pm$0.38 & 3.12$\pm$0.30 \\
\nodata & 2011 Dec 5 &	55900.3 & 2.514 & 4.62$\pm$0.63 & 	5.31$\pm$0.56 & 2.06$\pm$0.66 \\
\tableline
\multicolumn{7}{c}{{\it 1.5 m CTIO\/} RC}\\
\nodata & 2004 Jun 22 & 53178.1 & 1.169 & 10.51$\pm$1.15 & 8.56$\pm$0.84 & \nodata \\
\nodata & 2004 Nov 6 & 53315.4 & 1.236 & 9.02$\pm$0.84 & 7.39$\pm$0.63 & \nodata \\
\nodata & 2004 Nov 17 & 53326.3 & 1.242 & 10.84$\pm$1.31 & 8.15$\pm$1.33 & \nodata \\
\nodata & 2004 Nov 18 & 53327.4 & 1.242 & 10.37$\pm$1.85 & 7.94$\pm$1.34 & \nodata \\
\nodata & 2004 Nov 19 & 53328.3 & 1.243 & 10.47$\pm$1.49 & 8.06$\pm$1.08 & \nodata \\
\nodata & 2004 Nov 20 & 53329.3 & 1.243 & 10.53$\pm$1.96 & 8.33$\pm$1.43 & \nodata \\
\nodata & 2004 Nov 22 & 53331.3 & 1.244 & 10.32$\pm$1.00 & 7.71$\pm$0.73 & \nodata \\
\nodata & 2004 Dec 2 & 53341.3 & 1.249 & 10.79$\pm$0.65 & 8.62$\pm$0.48 & \nodata \\
\nodata & 2004 Dec 19 & 53358.4 & 1.258 & 10.64$\pm$1.05 & 8.49$\pm$0.78 & \nodata \\
\nodata & 2005 Jan 2 & 53372.3 & 1.265 & 10.64$\pm$1.36 & 7.56$\pm$0.96 & \nodata \\
\nodata & 2005 Jan 3 & 53373.4 & 1.265 & 10.70$\pm$1.36 & 8.01$\pm$1.00 & \nodata \\
\nodata & 2005 Jan 13 & 53383.4 & 1.270 & 11.11$\pm$1.21 & 8.05$\pm$0.87 & \nodata \\
\nodata & 2005 Jan 14 & 53384.2 & 1.271 & 11.00$\pm$1.35 & 8.02$\pm$0.97 & \nodata \\
\nodata & 2005 Jan 15 & 53385.3 & 1.271 & 10.03$\pm$1.12 & 7.29$\pm$0.80 & \nodata \\
\nodata & 2005 Jan 28 & 53398.3 & 1.277 & 10.92$\pm$1.17 & 8.02$\pm$0.84 & \nodata \\
\nodata & 2005 Jan 29 & 53399.3 & 1.278 & 10.31$\pm$1.38 & 7.65$\pm$1.01 & \nodata \\
\nodata & 2005 Feb 9 & 53410.4 & 1.283 & 10.90$\pm$1.12 & 7.98$\pm$0.81 & \nodata \\
\nodata & 2005 Feb 10 & 53411.4 & 1.284 & 10.81$\pm$0.79 & 7.97$\pm$0.57 & \nodata \\
\nodata & 2005 Mar 25 & 53454.1 & 1.305 & 10.56$\pm$1.01 & 8.55$\pm$0.75 & \nodata \\
\nodata & 2005 Mar 26 & 53455.2 & 1.306 & 10.79$\pm$0.75 & 8.42$\pm$0.55 & \nodata \\
\nodata & 2005 May 6 & 53496.1 & 1.326 & 10.83$\pm$1.36 & 7.77$\pm$0.98 & \nodata \\
\nodata & 2005 Jun 4 & 53525.1 & 1.340 & 10.71$\pm$0.94 & 7.97$\pm$0.68 & \nodata \\
\nodata & 2005 Jul 28 & 53580.0 & 1.367 & 9.87$\pm$1.44 & 7.61$\pm$1.05 & \nodata \\
\nodata & 2005 Jul 30 & 53582.0 & 1.368 & 9.93$\pm$0.97 & 7.90$\pm$0.72 & \nodata \\
\nodata & 2005 Nov 10 & 53684.4 & 1.419 & 10.63$\pm$1.22 & 7.37$\pm$0.87 & \nodata \\
\nodata & 2005 Nov 11 & 53685.4 & 1.419 & 10.95$\pm$1.30 & 7.95$\pm$0.93 & \nodata \\
\nodata & 2005 Nov 13 & 53687.3 & 1.420 & 10.87$\pm$1.33 & 7.82$\pm$0.95 & \nodata \\
\nodata & 2005 Nov 24 & 53698.4 & 1.426 & 10.33$\pm$1.46 & 7.08$\pm$1.05 & \nodata \\
\nodata & 2005 Nov 26 & 53700.4 & 1.427 & 10.78$\pm$1.73 & 7.71$\pm$1.24 & \nodata \\
\nodata & 2005 Nov 27 & 53701.3 & 1.427 & 10.90$\pm$1.88 & 7.79$\pm$1.35 & \nodata \\
\nodata & 2005 Dec 7 & 53711.3 & 1.432 & 11.18$\pm$1.59 & 7.93$\pm$1.15 & \nodata \\
\nodata & 2006 Jan 11 & 53746.3 & 1.449 & 9.46$\pm$1.85 & 6.92$\pm$1.34 & \nodata \\
\nodata & 2006 Jan 16 & 53751.2 & 1.452 & 11.36$\pm$1.46 & 7.92$\pm$1.04 & \nodata \\
\nodata & 2006 Jan 19 & 53754.3 & 1.453 & 10.84$\pm$1.55 & 7.36$\pm$1.09 & \nodata \\
\nodata & 2006 Jan 29 & 53764.2 & 1.458 & 11.73$\pm$1.57 & 8.12$\pm$1.14 & \nodata \\
\nodata & 2006 Jan 31 & 53766.2 & 1.459 & 11.63$\pm$1.79 & 8.28$\pm$1.28 & \nodata \\
\nodata & 2006 Mar 14 & 53808.1 & 1.480 & 11.77$\pm$1.32 & 8.04$\pm$0.94 & \nodata \\
\nodata & 2006 Mar 16 & 53810.1 & 1.481 & 11.34$\pm$1.13 & 7.94$\pm$0.82 & \nodata \\
\nodata & 2006 Mar 20 & 53814.2 & 1.483 & 9.85$\pm$1.43 & 6.92$\pm$1.06 & \nodata \\
\nodata & 2006 Apr 7 & 53832.1 & 1.492 & 11.29$\pm$1.80 & 7.70$\pm$1.27 & \nodata \\
\nodata & 2006 Jun 3 & 53889.9 & 1.520 & 12.56$\pm$1.41 & 8.39$\pm$0.99 & \nodata \\
\nodata & 2006 Aug 10 & 53958.0 & 1.554 & 12.36$\pm$1.31 & 8.54$\pm$0.93 & \nodata \\
\nodata & 2006 Aug 12 & 53960.0 & 1.555 & 12.04$\pm$1.45 & 8.30$\pm$1.03 & \nodata \\
\nodata & 2006 Oct 9 & 54017.4 & 1.583 & 12.05$\pm$1.39 & 8.88$\pm$1.00 & \nodata \\
\nodata & 2006 Oct 11 & 54019.4 & 1.584 & 11.57$\pm$1.15 & 8.60$\pm$0.83 & \nodata \\
\nodata & 2006 Oct 13 & 54021.4 & 1.585 & 8.10$\pm$2.10 & 5.82$\pm$1.52 & \nodata \\
\nodata & 2006 Oct 16 & 54024.4 & 1.587 & 11.70$\pm$1.19 & 8.73$\pm$0.86 & \nodata \\
\nodata & 2006 Dec 2 & 54071.3 & 1.610 & 11.73$\pm$1.22 & 8.97$\pm$0.88 & \nodata \\
\nodata & 2006 Dec 4 & 54073.3 & 1.611 & 11.84$\pm$1.02 & 8.91$\pm$0.74 & \nodata \\
\nodata & 2006 Dec 12 & 54081.4 & 1.615 & 8.84$\pm$0.93 & 6.85$\pm$0.68 & \nodata \\
\nodata & 2006 Dec 14 & 54083.4 & 1.616 & 8.55$\pm$0.97 & 6.34$\pm$0.71 & \nodata \\
\nodata & 2006 Dec 17 & 54086.2 & 1.618 & 11.65$\pm$0.83 & 8.63$\pm$0.61 & \nodata \\
\nodata & 2006 Dec 21 & 54090.3 & 1.620 & 11.20$\pm$0.98 & 8.79$\pm$0.71 & \nodata \\
\nodata & 2006 Dec 23 & 54092.3 & 1.621 & 9.09$\pm$1.16 & 7.37$\pm$0.85 & \nodata \\
\nodata & 2007 Jan 18 & 54118.3 & 1.633 & 11.38$\pm$1.03 & 8.87$\pm$0.76 & \nodata \\
\nodata & 2007 Jan 30 & 54130.4 & 1.639 & 11.52$\pm$0.93 & 8.77$\pm$0.66 & \nodata \\
\nodata & 2007 Feb 1 & 54132.2 & 1.640 & 10.71$\pm$0.82 & 8.57$\pm$0.61 & \nodata \\
\nodata & 2007 Feb 3 & 54134.3 & 1.641 & 9.46$\pm$0.77 & 7.95$\pm$0.57 & \nodata \\
\nodata & 2007 Feb 5 & 54136.1 & 1.642 & 11.21$\pm$0.81 & 8.77$\pm$0.59 & \nodata \\
\nodata & 2007 Feb 7 & 54138.1 & 1.643 & 7.49$\pm$0.95 & 6.23$\pm$0.72 & \nodata \\
\nodata & 2007 Feb 9 & 54140.2 & 1.644 & 11.18$\pm$0.96 & 8.68$\pm$0.70 & \nodata \\
\nodata & 2007 Feb 12 & 54143.2 & 1.646 & 11.39$\pm$0.86 & 8.90$\pm$0.63 & \nodata \\
\nodata & 2007 Feb 14 & 54145.1 & 1.647 & 11.00$\pm$0.99 & 8.84$\pm$0.73 & \nodata \\
\nodata & 2007 Feb 18 & 54149.1 & 1.649 & 7.52$\pm$0.77 & 6.09$\pm$0.57 & \nodata \\
\nodata & 2007 Mar 31 & 54190.1 & 1.669 & 9.48$\pm$0.58 & 7.95$\pm$0.43 & \nodata \\
\nodata & 2007 Apr 12 & 54202.1 & 1.675 & 10.39$\pm$0.85 & 8.66$\pm$0.64 & \nodata \\
\nodata & 2007 Jun 21 & 54272.9 & 1.710 & 9.54$\pm$1.05 & 8.12$\pm$0.79 & \nodata \\
\nodata & 2007 Jun 27 & 54279.0 & 1.713 & 9.92$\pm$0.29 & 8.27$\pm$0.22 & \nodata \\
\nodata & 2007 Jul 18 & 54300.0 & 1.723 & 10.23$\pm$0.74 & 8.22$\pm$0.55 & \nodata \\
\nodata & 2007 Jul 25 & 54306.0 & 1.726 & 9.36$\pm$0.54 & 7.74$\pm$0.41 & \nodata \\
\nodata & 2007 Jul 28 & 54310.0 & 1.728 & 10.61$\pm$0.86 & 8.37$\pm$0.63 & \nodata \\
\nodata & 2008 Feb 8 & 54504.3 & 1.824 & 8.90$\pm$0.45 & 7.85$\pm$0.34 & \nodata \\
\nodata & 2008 Feb 13 & 54509.3 & 1.827 & 8.66$\pm$0.58 & 7.30$\pm$0.44 & \nodata \\
\nodata & 2008 Feb 19 & 54515.3 & 1.830 & 9.10$\pm$0.25 & 8.49$\pm$0.19 & \nodata \\
\nodata & 2008 Feb 26 & 54522.3 & 1.833 & 9.48$\pm$0.37 & 7.80$\pm$0.28 & \nodata \\
\nodata & 2008 Feb 27 & 54523.2 & 1.833 & 8.20$\pm$0.87 & 6.99$\pm$0.65 & \nodata \\
\nodata & 2008 Mar 1 & 54526.2 & 1.835 & 10.06$\pm$0.69 & 8.31$\pm$0.52 & \nodata \\
\nodata & 2008 Mar 3 & 54528.2 & 1.836 & 9.62$\pm$1.01 & 7.81$\pm$0.74 & \nodata \\
\nodata & 2008 Mar 7 & 54532.3 & 1.838 & 8.28$\pm$0.65 & 7.16$\pm$0.49 & \nodata \\
\nodata & 2008 Mar 9 & 54534.2 & 1.839 & 8.11$\pm$0.28 & 7.19$\pm$0.21 & \nodata \\
\nodata & 2008 Mar 15 & 54540.2 & 1.842 & 9.30$\pm$0.48 & 7.92$\pm$0.36 & \nodata \\
\nodata & 2008 Mar 25 & 54550.2 & 1.847 & 7.32$\pm$0.11 & 6.43$\pm$0.09 & \nodata \\
\nodata & 2008 Mar 30 & 54555.1 & 1.849 & 9.56$\pm$0.52 & 8.12$\pm$0.39 & \nodata \\
\nodata & 2008 Apr 3 & 54559.1 & 1.851 & 6.73$\pm$0.47 & 6.37$\pm$0.36 & \nodata \\
\nodata & 2008 Apr 14 & 54570.1 & 1.857 & 9.84$\pm$0.61 & 8.41$\pm$0.46 & \nodata \\
\nodata & 2008 Apr 16 & 54572.1 & 1.858 & 9.12$\pm$0.61 & 7.95$\pm$0.46 & \nodata \\
\nodata & 2008 Apr 22 & 54578.1 & 1.861 & 7.40$\pm$0.18 & 6.55$\pm$0.14 & \nodata \\
\nodata & 2008 May 4 & 54590.1 & 1.867 & 7.86$\pm$0.47 & 7.54$\pm$0.36 & \nodata \\
\nodata & 2008 May 11 & 54597.1 & 1.870 & 8.14$\pm$0.18 & 7.26$\pm$0.14 & \nodata \\
\nodata & 2008 May 16 & 54602.1 & 1.873 & 9.54$\pm$0.47 & 8.44$\pm$0.35 & \nodata \\
\nodata & 2008 Jun 22 & 54639.0 & 1.891 & 8.56$\pm$0.24 & 7.72$\pm$0.18 & \nodata \\
\nodata & 2008 Jun 27 & 54645.0 & 1.894 & 9.07$\pm$0.41 & 8.04$\pm$0.31 & \nodata \\
\nodata & 2008 Jul 3 & 54651.0 & 1.897 & 9.24$\pm$0.24 & 8.33$\pm$0.18 & \nodata \\
\nodata & 2008 Jul 10 & 54657.0 & 1.900 & 9.63$\pm$0.02 & 8.62$\pm$0.01 & \nodata \\
\nodata & 2008 Jul 13 & 54661.0 & 1.902 & 8.72$\pm$0.61 & 7.75$\pm$0.46 & \nodata \\
\nodata & 2008 Nov 6 & 54776.3 & 1.959 & 8.78$\pm$0.27 & 8.32$\pm$0.20 & \nodata \\
\nodata & 2008 Dec 4 & 54804.3 & 1.972 & 9.17$\pm$0.26 & 8.52$\pm$0.20 & \nodata \\
\nodata & 2008 Dec 8 & 54808.3 & 1.974 & 8.81$\pm$0.47 & 8.36$\pm$0.36 & \nodata \\
\nodata & 2008 Dec 15 & 54815.3 & 1.978 & 9.52$\pm$0.15 & 9.06$\pm$0.12 & \nodata \\
\nodata & 2008 Dec 17 & 54817.3 & 1.979 & 9.20$\pm$0.13 & 9.01$\pm$0.09 & \nodata \\
\nodata & 2008 Dec 22 & 54822.4 & 1.981 & 8.78$\pm$0.12 & 8.66$\pm$0.09 & \nodata \\
\nodata & 2008 Dec 28 & 54828.4 & 1.984 & 8.30$\pm$0.24 & 8.36$\pm$0.18 & \nodata \\
\nodata & 2009 Mar 8 & 54898.1 & 2.019 & 6.37$\pm$0.28 & 6.00$\pm$0.21 & \nodata \\
\nodata & 2010 Jan 10 & 55206.3 & 2.171 & 5.30$\pm$0.25 & 4.58$\pm$0.19 & \nodata \\
\nodata & 2010 Jan 21 & 55217.3 & 2.177 & 5.72$\pm$0.11 & 4.87$\pm$0.09 & \nodata \\
\nodata & 2010 Apr 25 & 55312.0 & 2.223 & 5.96$\pm$0.45 & 4.46$\pm$0.33 & \nodata \\
\nodata & 2010 Aug 2 & 55411.0 & 2.272 & 5.65$\pm$1.44 & 5.07$\pm$1.09 & \nodata \\
\nodata & 2010 Oct 30 & 55499.4 & 2.316 & 5.81$\pm$0.36 & 4.47$\pm$0.27 & \nodata \\
\nodata & 2010 Nov 11 & 55511.4 & 2.322 & 4.41$\pm$3.34 & 3.57$\pm$2.55 & \nodata \\
\nodata & 2010 Nov 16 & 55516.3 & 2.324 & 5.66$\pm$0.74 & 4.48$\pm$0.56 & \nodata \\
\nodata & 2011 Mar 6 & 55626.2 & 2.379 & 5.90$\pm$0.64 & 4.47$\pm$0.48 & \nodata \\
\nodata & 2011 Apr 17 & 55668.1 & 2.399 & 4.77$\pm$0.37 & 4.40$\pm$0.28 & \nodata \\
\nodata & 2011 Nov 30 & 55895.3 & 2.512 & 4.29$\pm$0.01 & 4.14$\pm$0.01 & \nodata \\
\nodata & 2011 Dec 16 & 55911.3 & 2.520 & 4.41$\pm$0.34 & 4.10$\pm$0.27 & \nodata \\
\nodata & 2011 Dec 19 & 55914.3 & 2.521 & 4.83$\pm$0.46 & 4.64$\pm$0.27 & \nodata \\
\nodata & 2011 Dec 21 & 55916.2 & 2.522 & 4.57$\pm$0.35 & 3.69$\pm$0.27 & \nodata \\
\nodata & 2011 Dec 28 & 55923.2 & 2.526 & 5.44$\pm$0.21 & 4.28$\pm$0.16 & \nodata \\
\nodata & 2012 Jan 5 & 55931.2 & 2.530 & 5.14$\pm$0.48 & 4.54$\pm$0.27 & \nodata \\
\nodata & 2012 Jan 17 & 55943.4 & 2.536 & 4.76$\pm$0.19 & 4.43$\pm$0.14 & \nodata \\
\enddata
\tablecomments{}
	\tablenotetext{a}{Mainly \ion{Fe}{2}/\ion{Cr}{2} blends.}
	\tablenotetext{b}{As listed on the Eta Carinae Treasury Project site at \protect{http://etacar.umn.edu/}.}  
	\tablenotetext{c}{Integration range as description, continuum was set at $\lambda\lambda$4600--4610 and $\lambda\lambda$4740--4744 \AA.}
\end{deluxetable}

\begin{deluxetable}{lccccccc}
\tabletypesize{\scriptsize}
\tablecaption{Equivalent Widths of H$\alpha$ and H$\delta$ (1998--2012)\label{tab:table2}}
\tablewidth{0pt}
\tablehead{
\colhead{Name\tablenotemark{a}} &
\colhead{Date} &
\colhead{MJD}   &
\colhead{Phase} &
\colhead{EW$_{H\alpha\tablenotemark{b}}^{Star}$} &
\colhead{EW$_{H\alpha\tablenotemark{b}}^{FOS4}$}  &
\colhead{EW$_{H\delta\tablenotemark{c}}^{Star}$} &
\colhead{EW$_{H\delta\tablenotemark{c}}^{FOS4}$}    \\
\colhead{} &
\colhead{(UT)} &
\colhead{}   &
\colhead{} &
\colhead{(\AA)} &
\colhead{(\AA)} &
\colhead{(\AA)} &
\colhead{(\AA)}
}
   \startdata
\multicolumn{8}{c}{{\it HST\/} STIS}\\
c800 & 1998	Jan 1 &	50814.1 &	0.000 &	847.67$\pm$11.48 & \nodata & \nodata & \nodata\\
c821 & 1998 Mar 19 &	50891.5 &	0.038 &	824.19$\pm$13.36 & \nodata & 30.79$\pm$1.01   & \nodata \\
c890 & 1998	Nov 25 &	51142.2 &	0.162 &	887.71$\pm$15.39 & \nodata  & \nodata & \nodata \\
c914 &1999	Feb 21 &	51230.5 & 0.206 &	873.64$\pm$9.57 & \nodata & 32.69$\pm$0.01 & \nodata  \\
cA20 &2000 Mar 13 &	51616.5 &	0.397 &	814.91$\pm$48.81 & \nodata & \nodata & \nodata \\
cA22 &2000 Mar 20 &	51623.8 &	0.400 &	836.04$\pm$14.31 & \nodata & 31.79$\pm$0.38 & \nodata  \\
cA22 &2000 Mar 21 &	51624.5 &	0.401 &	791.12$\pm$35.31 & \nodata  & \nodata & \nodata \\
cB29 &2001 Apr 17 &	52016.8 &	0.595 &	763.89$\pm$7.94 & \nodata & 32.07$\pm$0.46 & \nodata \\
cB75 &	2001 Oct 1 &	52183.2 &	0.677 &	828.76$\pm$11.14 & \nodata & 32.91$\pm$0.02 & \nodata \\
cB90 &2001	Nov 27 &	52240.1 &	0.705 &	859.09$\pm$11.27 & \nodata  & \nodata & \nodata \\
cC05 & 2002	Jan 19 &	52294.1 &	0.732 &	896.02$\pm$20.90 & \nodata & 35.45$\pm$0.22 & \nodata \\
cC51 &2002 Jul 4 &	52459.6 &	0.813 &	886.67$\pm$11.25 & \nodata & 35.31$\pm$0.35 & \nodata \\
cC96 &2002	Dec 16 &	52624.1 &	0.895 &	944.16$\pm$21.31 & \nodata & \nodata & \nodata \\
cD12 &2003 	Feb 12 &	52683.0 &	0.924 &	936.87$\pm$18.63 & \nodata & 33.23$\pm$0.36 & \nodata \\
cD24 & 2003	Mar 29 &	52727.2 &	0.946 &	851.55$\pm$14.13 & \nodata & 29.75$\pm$0.31 & \nodata \\
cD34 &2003 	May 5 &	52764.3 &	0.964 &	741.85$\pm$11.66 & \nodata & 27.81$\pm$0.05 & \nodata \\
cD37 &2003 	May 17 &	52776.4 &	0.970 &	716.10$\pm$10.06 & \nodata & \nodata & \nodata  \\
cD37 &2003 	May 19 &	52778.5 &	0.971 &	716.74$\pm$13.73& \nodata & 27.74$\pm$0.01 & \nodata \\
cD41 &2003 	May 26 &	52785.8 &	0.975 &	691.44$\pm$10.68 & \nodata & \nodata & \nodata \\
cD41 &2003 	Jun 1 &	52791.6 &	0.978 &	672.51$\pm$13.12 & \nodata & 27.23$\pm$0.04 & \nodata \\
cD47 &2003 	Jun 22 &	52812.2 &	0.988 &	569.05$\pm$5.52 & \nodata & \nodata & \nodata  \\
cD47 &2003 	Jun 23 &	52813.7 &	0.988 &	547.58$\pm$8.78 & \nodata & 24.24$\pm$0.77 & \nodata \\
cD51 &2003 	Jul 5 &	52825.2 &	0.994 &	547.88$\pm$21.15 & \nodata & 23.26$\pm$1.87 & \nodata \\
cD58 &	2003 Jul 29 &	52849.6 &	1.006 & 527.93$\pm$10.37 & \nodata & \nodata & \nodata  \\
cD58 &	2003 Jul 31 &	52852.1 &	1.007 & 524.79$\pm$15.08 & \nodata & 22.12$\pm$1.63 & \nodata   \\
cD72 &	2003 Sep 22 &	52904.4 &	1.033 &	599.94$\pm$7.27 & \nodata & 28.20$\pm$1.63 & \nodata   \\
cD88 &	2003 Nov 17 &	52960.6 &	1.061 &	682.24$\pm$9.31& \nodata & 30.35$\pm$0.93 & \nodata  \\
cE18 &2004	Mar 7 &	53071.3 &1.116 &	802.45$\pm$7.17 & \nodata & 28.06$\pm$0.30  & \nodata \\
cJ63 & 2009	Aug 18 &	55062.0 & 2.100 &	468.53$\pm$3.62 & \nodata & \nodata & \nodata  \\
cK16 & 2010	Mar 3 &	55258.6 &	2.197 &	495.75$\pm$5.86 & \nodata & \nodata & \nodata  \\
cK63 & 2010	Aug 20 &	55428.3 &	2.281 &	493.60$\pm$7.95 & \nodata  & \nodata & \nodata  \\
\tableline
\multicolumn{8}{c}{{\it VLT\/} UVES}\\
uC93 &  2002 Dec 7 & 52615.3 & 0.890 & 917.37$\pm$30.17 & \nodata & 24.19$\pm$0.23 & \nodata\\
uC93 & 2002 Dec 8 &  52616.3 & 0.891 & \nodata & 557.92$\pm$0.97  & \nodata & \nodata \\
uC95 & 2002 Dec 12 & 52620.3 & 0.893 & 926.76$\pm$30.50 & \nodata & 24.91$\pm$0.31 & \nodata \\
uC98 & 2002 Dec 26 & 52634.3& 0.900 & \nodata & 558.57$\pm$5.43  & \nodata & 18.20$\pm$1.06 \\
uD00 & 2002 Dec 31 & 52639.4 & 0.902 & \nodata & 558.60$\pm$4.95  & \nodata & 18.73$\pm$0.95 \\
uD00 & 2003 Jan 3 & 52642.3 & 0.904 & \nodata & 559.62$\pm$3.76  & \nodata & 18.98$\pm$1.53 \\
uD05 & 2003 Jan 19 & 52658.3 & 0.912 & \nodata & 590.53$\pm$5.76  & \nodata & \nodata \\
uD05 & 2003 Jan 23 & 52662.4 & 0.914 & \nodata & 610.34$\pm$0.21  & \nodata & 19.58$\pm$0.67 \\
uD09 & 2003 Feb 4 &52674.4 & 0.920 & \nodata & 585.55$\pm$5.72  & \nodata & 19.07$\pm$0.87 \\
uD12 & 2003 Feb 14 & 52684.1 & 0.924 & 932.59$\pm$27.77 & 577.99$\pm$0.37 & 24.84$\pm$0.28 & 18.29$\pm$0.78\\
uD15 & 2003 Feb 25 & 52695.3 & 0.930 &  \nodata & 590.86$\pm$9.45 & \nodata & 19.23$\pm$0.71  \\
uD18 & 2003 Mar 7 & 52705.3 & 0.935 &  \nodata & 569.04$\pm$0.26  & \nodata & \nodata \\
uD18 & 2003 Mar 12 & 52710.0 & 0.937 &  \nodata & 571.14$\pm$8.47 & \nodata & 18.48$\pm$0.89  \\
uD33 & 2003 Apr 30 & 52759.1 & 0.962 &  \nodata & 506.91$\pm$1.56 & \nodata & 16.49$\pm$0.98  \\
uD33 & 2003 May 5 & 52765.0 & 0.964 &  \nodata & 502.40$\pm$1.31  & \nodata & 17.03$\pm$0.55 \\
uD36 & 2003 May 12 & 52771.2 & 0.967 &  \nodata & 503.97$\pm$1.44 & \nodata & 16.87$\pm$0.54  \\
uD40 & 2003 May 29 & 52788.1 & 0.976 & 715.23$\pm$25.60 & 483.87$\pm$0.97 & 20.54$\pm$0.05 & 16.59$\pm$0.92\\
uD42 & 2003 Jun 4 & 52794.0 & 0.979 & 700.51$\pm$22.36 & 477.21$\pm$0.60 & 20.20$\pm$0.07 & 16.90$\pm$0.89\\
uD42 & 2003 Jun 8 & 52798.01 & 0.981 & \nodata & 481.12$\pm$0.13 & \nodata & 16.85$\pm$0.86  \\
uD45 & 2003 Jun 12 & 52803.0 & 0.983 & \nodata & 481.89$\pm$1.94 & \nodata & 16.83$\pm$0.85  \\
uD45 & 2003 Jun 17 & 52808.0 & 0.986 & \nodata & 475.60$\pm$0.51 & \nodata & 16.42$\pm$1.02  \\
uD47 & 2003 Jun 22 & 52813.0 & 0.988 & \nodata & 467.47$\pm$0.99 & \nodata & 16.29$\pm$0.64  \\
uD49 & 2003 Jun 30 & 52821.0 & 0.992 & \nodata & 449.68$\pm$1.29 & \nodata & 15.86$\pm$1.13  \\
uD51 & 2003 Jul 5 & 52825.0 & 0.994 & 548.32$\pm$17.48 & 446.94$\pm$5.55 & 16.02$\pm$1.66 & 16.20$\pm$1.07\\
uD51 & 2003 Jul 9 & 52830.0 & 0.997 & \nodata & 447.45$\pm$0.72  & \nodata & 15.72$\pm$1.15 \\
uD54 & 2003 Jul 16 & 52836.0 & 1.000 & \nodata & 457.06$\pm$3.25 & \nodata & 15.37$\pm$1.10  \\
uD54 & 2003 Jul 20 & 52841.0 & 1.002 & \nodata & 430.21$\pm$5.55 & \nodata & 14.95$\pm$1.46  \\
uD57 & 2003 Jul 26 & 52847.0 & 1.005 & \nodata & 436.51$\pm$7.31 & \nodata & \nodata  \\
uD57 & 2003 Jul 27 & 52848.0 & 1.005 & \nodata & \nodata  & \nodata & 15.49$\pm$1.28\\
uD57 & 2003 Jul 31 & 52852.0 & 1.007 & \nodata & 439.33$\pm$6.84 & \nodata & 15.20$\pm$1.19 \\
uD90 & 2003 Nov 25 & 52968.3 & 1.065 & \nodata & 487.43$\pm$4.00 & \nodata &20.25$\pm$1.34  \\
uD96 & 2003 Dec 17 & 52990.3 & 1.076 & \nodata & 508.95$\pm$3.15 & \nodata & 20.13$\pm$1.48  \\
uE00 & 2004 Jan 2 & 53006.3 & 1.084 & \nodata & 546.26$\pm$6.21  & \nodata & 20.69$\pm$1.31 \\
uE07 & 2004 Jan 25 & 53029.3 & 1.095 & \nodata & 567.36$\pm$2.95 & \nodata & 21.09$\pm$1.38  \\
uE14 & 2004 Feb 20 & 53055.1 & 1.108 & 855.81$\pm$24.21 & 604.02$\pm$4.24 & 27.20$\pm$0.48 & 21.52$\pm$1.41\\
uE19 & 2004 Mar11 & 53075.1& 1.118 & \nodata & 644.83$\pm$14.56 & \nodata & 21.13$\pm$1.02  \\
uE94 & 2004 Dec 10 & 53349.4 & 1.253 & \nodata & 540.65$\pm$2.77 & \nodata & 19.46$\pm$0.81  \\
uF05 & 2005 Jan 19 & 53389.2 & 1.273 & \nodata & 538.42$\pm$1.78 & \nodata & 19.51$\pm$0.82  \\
uF12 & 2005 Feb 12 & 53413.4 & 1.285 & 693.55$\pm$16.05 & \nodata & 25.63$\pm$0.50 & \nodata \\
uF17 & 2005 Mar 2 & 53431.3 & 1.294 & \nodata & 533.66$\pm$2.34 & \nodata & 19.32$\pm$0.74  \\
uF21 & 2005 Mar 19 & 53448.1 & 1.302 & 724.45$\pm$17.71& \nodata & 26.91$\pm$0.71 & \nodata \\  
uG27 & 2006 Apr 9 & 53834.1 & 1.493 & 704.31$\pm$18.31 & \nodata & 26.25$\pm$0.55 &  \nodata \\
uG36 & 2006 May 11 & 53866.01 & 1.509 & \nodata & 535.39$\pm$2.66 & \nodata & 19.14$\pm$1.00  \\
uG43 & 2006 Jun 8 & 53894.0 & 1.522 & 741.38$\pm$14.13 &\nodata & 26.88$\pm$0.46 & \nodata\\
uG48 & 2006 Jun 26 & 53912.04 & 1.531 & \nodata & 580.11$\pm$3.58 & \nodata & 21.26$\pm$0.88  \\
uI02 & 2008 Jan 10 & 54475.3 & 1.810 & 806.36$\pm$17.96 & 619.01$\pm$2.24 & 26.99$\pm$1.11 & \nodata \\
uI13 & 2008 Feb 17 & 54513.3 & 1.829 & 784.58$\pm$14.67 & 580.60$\pm$1.50 & 27.00$\pm$1.01 & 19.23$\pm$0.55\\
uI14 & 2008 Feb 21 & 54517.2 & 1.831 & 814.82$\pm$82.18 & \nodata  & \nodata & \nodata \\
uI19 & 2008 Mar 10 & 54535.3 & 1.839 & 769.93$\pm$16.99 & 615.62$\pm$1.37 & 26.58$\pm$0.65 & \nodata \\
uI24 & 2008 Mar 29 & 54554.3 & 1.849 & 822.03$\pm$18.81 & 600.54$\pm$1.09 & 28.51$\pm$0.97 & 20.75$\pm$0.50\\
uI28 & 2008 Apr 11 & 54567.0 & 1.855 & 827.23$\pm$19.37 & 618.10$\pm$1.04 & 27.14$\pm$0.70 & 20.78$\pm$0.56\\
uI32 & 2008 Apr 27 & 54583.0 & 1.863 & 780.82$\pm$16.50 & 591.73$\pm$9.99 & 26.36$\pm$0.60 & 19.43$\pm$0.61\\
uI36 & 2008 May 12 & 54599.0 & 1.871 & 794.97$\pm$14.78 & 599.38$\pm$0.58 & 26.40$\pm$0.68 & 19.20$\pm$0.66\\
uI41 & 2008 May 28 & 54615.0 & 1.879 & \nodata & \nodata & 25.37$\pm$0.45 & \nodata \\
uI41 & 2008 May 30 & 54616.0 & 1.879 & 819.87$\pm$17.26 & 629.67$\pm$2.72 & 25.49$\pm$0.53 & 20.12$\pm$0.17\\
uI41 & 2008 May 31 & 54617.1 & 1.880 & 825.33$\pm$20.18 & 631.87$\pm$2.33 & 25.50$\pm$0.47 & 19.47$\pm$0.58\\
uI44 & 2008 Jun 11 & 54629.0 & 1.886 & 809.55$\pm$17.82 & 608.96$\pm$0.47 & 25.04$\pm$0.37 & 19.11$\pm$0.67\\
uI52 & 2008 Jul 9 & 54656.0 & 1.899 & 794.27$\pm$19.01 & 608.44$\pm$1.61 & 24.14$\pm$0.47 & 19.27$\pm$0.20\\
uI52 & 2008 Jul 10 & 54657.04 & 1.900 & \nodata & 599.91$\pm$.78 & \nodata & 18.43$\pm$0.54  \\
uJ03 & 2009 Jan 10 & 54841.4 & 1.991 & 481.47$\pm$12.15 & \nodata & 16.38$\pm$0.68 & \nodata \\
uJ07 & 2009 Jan 25 & 54856.19 & 1.998 & \nodata & 491.19$\pm$4.23  & \nodata & 17.11$\pm$1.08  \\
uJ10 & 2009 Feb 5 & 54867.2 & 2.004 & 494.66$\pm$15.37 & \nodata & 15.77$\pm$1.33 & \nodata \\
uJ13 & 2009 Feb 19 & 54881.0 & 2.010 & \nodata & 640.94$\pm$13.69  & \nodata & \nodata   \\
uJ14 & 2009 Feb 20 & 54882.2 & 2.011 & \nodata & 497.64$\pm$7.05  & \nodata & 17.09$\pm$0.80   \\
uJ25 & 2009 Apr 2 & 54923.2 & 2.031 & 459.72$\pm$9.01 & \nodata & 15.51$\pm$0.23 & 17.05$\pm$0.96\\
uJ31 & 2009 Apr 25 & 54946.1 & 2.043 & 485.44$\pm$8.58 & \nodata & 16.60$\pm$0.30 & 18.06$\pm$0.94 \\
uJ46 & 2009 Jun 17 & 54999.1 & 2.069 & \nodata & \nodata & 16.96$\pm$0.52 & 18.64$\pm$0.94\\
uJ50 & 2009 Jun 30 & 55013.0 & 2.076 & \nodata & \nodata & 15.97$\pm$0.36 & 18.78$\pm$0.49\\
uJ50 & 2009 Jul 1 & 55014.0 & 2.076 & \nodata & \nodata & 15.52$\pm$0.37 & 18.85$\pm$0.81\\
uJ50 & 2009 Jul 2 & 55014.0 & 2.077 & \nodata & \nodata & 16.50$\pm$0.43 & 17.60$\pm$0.32\\
uJ51 & 2009 Jul 5 & 55018.0 & 2.078 & \nodata & \nodata & 14.91$\pm$0.27 & 18.58$\pm$0.70\\
\tableline
\multicolumn{8}{c}{{\it Gemini\/} GMOS}\\
gH45 & 2007 Jun 16 &  54267.1 &1.707 &	\nodata & \nodata & 25.97$\pm$0.48 & \nodata \\
gH45 & 2007 Jun 17 & 54269.0 & 1.708 & 	\nodata & \nodata & 26.3642875 0.31 & \nodata \\
gH49 &	 2007 Jun 29 &		54281.0 &		1.714 & \nodata & 	604.36$\pm$0.15 \nodata & \nodata \\
gI11 & 2008 Feb 11 & 54507.4 & \nodata \nodata & 26.14$\pm$0.31 & 17.00$\pm$1.35 \\
gI11 &	2008 Feb 13 &		54509.2 &		1.827 & \nodata & 	557.30$\pm$2.09 \nodata & \nodata \\
gI50 &	2008 Jul 4 &		54652.0 &		1.897 & \nodata & 	564.62$\pm$1.417  \nodata & \nodata \\
gI54 & 2008 Jul 17 & 54664.0 & 1.903 & \nodata & \nodata & 24.47$\pm$0.28 & \nodata \\
gI54 & 2008 Jul 18 & 54665.0 & 1.904 & \nodata & \nodata & 24.54$\pm$0.30 &  18.34$\pm$0.65 \\
gI85 &	2008 Nov 8 &		54778.3 &		1.960 & \nodata & 	581.56$\pm$54.89 & \nodata & 19.23$\pm$2.27 \\
gI90 &	2008 Nov 27 &		54797.3 &		1.969 & \nodata & 	508.12$\pm$17.94 & 23.38$\pm$0.65 & 16.37$\pm$1.18 \\
gI96 &	2008 Dec 18 &		54818.3 &		1.979 & \nodata & 	500.06$\pm$6.55 & 21.75$\pm$0.85 & 17.30$\pm$0.67 \\
gI98 &	2008 Dec 25 &		54825.3 &		1.983 & \nodata & 	491.51$\pm$10.14 & \nodata & 17.41$\pm$0.59\\
gI99 &	2008 Dec 31 &		54831.3 &		1.986 & \nodata & 	507.55$\pm$55.31  & 20.12$\pm$0.24 & 16.48$\pm$0.63 \\
gJ01 &	2009 Jan 4 &		54835.3 &		1.988 & \nodata & 	479.13$\pm$7.79 & 19.31$\pm$0.10 & 16.87$\pm$0.71 \\
gJ02 &	2009 Jan 9 &		54840.2 &		1.990 & \nodata & 	474.08$\pm$11.29 & 17.55$\pm$0.53 & 16.18$\pm$0.91 \\
gJ03 &	2009 Jan 12 &		54843.3 &		1.992 & \nodata & 	472.01$\pm$8.21 & 16.06$\pm$1.09 & 16.84$\pm$0.99\\
gJ04 &	2009 Jan 15 &		54846.2 &		1.993 & \nodata & 	453.95$\pm$26.46 & 15.97$\pm$1.28 & 15.97$\pm$0.95 \\
gJ05 &	2009 Jan 21 &		54852.3 &		1.996 & \nodata & 	460.93$\pm$7.44 & 17.73$\pm$1.24 & 16.44$\pm$1.23 \\
gJ06 &	2009 Jan 24 &		54855.3 &		1.998 & \nodata & 	466.87$\pm$4.80 & 18.24$\pm$1.21 & 16.40$\pm$1.20 \\
gJ07 &	2009 Jan 29 &		54860.3 &		2.000 & \nodata & 	457.18$\pm$10.59 &  17.86$\pm$1.17 & 15.25$\pm$1.11 \\
gJ09 &	2009 Feb 5 &		54867.2 &		2.004 & \nodata & 	460.21$\pm$13.58 & 16.14$\pm$1.15 & 15.48$\pm$0.81 \\
gJ13 &	2009 Feb 19 &		54881.2 &		2.010 & \nodata & 	464.74$\pm$8.33 & 15.32$\pm$1.03 & 16.12$\pm$1.08 \\
gJ20 & 2009 Mar 17 & 54907.3 & 2.023 & \nodata & \nodata & 16.19$\pm$0.23 & \nodata \\
gJ32 &	2009 Apr 28 &		54949.1 &		2.044 & \nodata & 	483.66$\pm$7.57 & 17.48$\pm$0.23 & 17.6$\pm$1.19 \\
gJ56 &	2009 Jul 23 &		55036.0 &		2.087 & \nodata & 	489.75$\pm$15.67 & 17.99$\pm$0.10 &  \nodata \\
gK02	 & 2010	Jan 8 &		55204.4 &		2.170 & \nodata & 	527.29$\pm$1.43 & 18.27$\pm$0.21 & \nodata \\
gK05 &	 2010 Jan 20 &	55216.3 &		2.176 & \nodata & 	523.09$\pm$0.18 \nodata & \nodata \\
\tableline
\multicolumn{8}{c}{{\it Magellan II\/} MIKE}\\
\nodata & 2010 Jun 4 & 55351.9 & 2.243 & \nodata & \nodata & 18.76$\pm$0.08 & \nodata \\
\nodata & 2010 Jun 5 & 55353.5 & 2.244 & 515.41$\pm$15.10 & \nodata & 18.89$\pm$0.30 & \nodata \\
\tableline  
\multicolumn{8}{c}{{\it Ir\'{e}n\'{e}e du Pont\/} B\&C}\\
\nodata &	2011 Feb 25 & 55629.2 &	2.381 & \nodata & \nodata & 21.57$\pm$0.37 & 19.10$\pm$0.52 \\
\nodata &	2011 Feb 26 & 55630.7 &	2.381 & \nodata & 499.02$\pm$97.38 & \nodata & \nodata \\
\nodata &	2011 Jun 8 & 55721.0 &	2.426 &	\nodata & \nodata & 19.30$\pm$0.66 & 18.29$\pm$1.84 \\
\nodata &	2011 Jun 9 & 55722.0 &	2.426 &	604.56$\pm$13.51 & 652.31$\pm$22.73 & \nodata & \nodata \\
\nodata &2011 Dec 5 & 55900.3 & 	2.514 & \nodata & \nodata &  	19.47$\pm$1.22 & 17.26$\pm$1.53 \\
\nodata & 2011 Dec 6 &	55901.3 & 2.515 & 588.08$\pm$41.49 & 525.15$\pm$20.44 & \nodata & \nodata\\
\tableline
\multicolumn{8}{c}{{\it 1.5 m CTIO\/} RC}\\
\nodata & 2004 Jun 22 & 53178.1 & 1.169 & \nodata & \nodata  & 24.08$\pm$0.42 & \nodata \\ 
\nodata & 2004 Nov 6 & 53315.4 & 1.236 & \nodata & \nodata  & 23.91$\pm$0.91 & \nodata \\ 
\nodata & 2004 Nov 17 & 53326.3 & 1.242 & \nodata & \nodata  & 24.82$\pm$0.84 & \nodata \\ 
\nodata & 2004 Nov 18 & 53327.4 & 1.242 & \nodata & \nodata  & 24.95$\pm$0.66 & \nodata \\ 
\nodata & 2004 Nov 19 & 53328.3 & 1.243 & \nodata & \nodata  & 24.57$\pm$0.73 & \nodata \\ 
\nodata & 2004 Nov 20 & 53329.3 & 1.243 & \nodata & \nodata  & 25.08$\pm$0.91 & \nodata \\ 
\nodata & 2004 Nov 22 & 53331.3 & 1.244 & \nodata & \nodata  & 24.60$\pm$0.93 & \nodata \\ 
\nodata & 2004 Dec 2 & 53341.3 & 1.249 & \nodata & \nodata  & 26.89$\pm$0.98 & \nodata \\ 
\nodata & 2004 Dec 19 & 53358.4 & 1.258 & \nodata & \nodata  & 26.57$\pm$1.34 & \nodata \\ 
\nodata & 2005 Jan 2 & 53372.3 & 1.265 & \nodata & \nodata  & 25.37$\pm$0.81 & \nodata \\ 
\nodata & 2005 Jan 3 & 53373.4 & 1.265 & \nodata & \nodata  & 25.39$\pm$1.42 & \nodata \\ 
\nodata & 2005 Jan 13 & 53383.4 & 1.270 & \nodata & \nodata  & 25.19$\pm$0.80 & \nodata \\ 
\nodata & 2005 Jan 14 & 53384.2 & 1.271 & \nodata & \nodata  & 25.09$\pm$0.92 & \nodata \\ 
\nodata & 2005 Jan 15 & 53385.3 & 1.271 & \nodata & \nodata  & 24.29$\pm$0.67 & \nodata \\ 
\nodata &  2005 Jan 17 & 	53387.3 &	    1.272 & 690.88$\pm$16.21 & \nodata & \nodata & \nodata \\
\nodata & 2005 Jan 28 & 53398.3 & 1.277 & \nodata & \nodata  & 24.35$\pm$0.82 & \nodata \\ 
\nodata & 2005 Jan 29 & 53399.3 & 1.278 & \nodata & \nodata  & 24.06$\pm$0.82 & \nodata \\ 
\nodata &  2005 Jan 31 & 	53401.3 &	    1.279 & 680.26$\pm$17.57 & \nodata & \nodata & \nodata  \\
\nodata & 2005 Feb 9 & 53410.4 & 1.283 & \nodata & \nodata  & 25.51$\pm$0.89 & \nodata \\ 
\nodata & 2005 Feb 10 & 53411.4 & 1.284 & \nodata & \nodata  & 24.87$\pm$0.94 & \nodata \\ 
\nodata & 2005 Mar 25 & 53454.1 & 1.305 & \nodata & \nodata  & 28.19$\pm$1.18 & \nodata \\ 
\nodata & 2005 Mar 26 & 53455.2 & 1.306 & \nodata & \nodata  & 25.58$\pm$0.40 & \nodata \\ 
\nodata & 2005 May 6 & 53496.1 & 1.326 & \nodata & \nodata  & 27.11$\pm$0.93 & \nodata \\ 
\nodata & 2005 Jun 4 & 53525.1 & 1.340 & \nodata & \nodata  & 27.06$\pm$0.91 & \nodata \\ 
\nodata & 2005 Jul 28 & 53580.0 & 1.367 & \nodata & \nodata  & 26.22$\pm$0.81 & \nodata \\ 
\nodata & 2005 Jul 30 & 53582.0 & 1.368 & \nodata & \nodata  & 26.00$\pm$0.80 & \nodata \\ 
\nodata & 2005 Nov 10 & 53684.4 & 1.419 & \nodata & \nodata  & 25.15$\pm$0.82 & \nodata \\ 
\nodata & 2005 Nov 11 & 53685.4 & 1.419 & \nodata & \nodata  & 25.77$\pm$1.02 & \nodata \\ 
\nodata & 2005 Nov 13 & 53687.3 & 1.420 & \nodata & \nodata  & 24.97$\pm$0.47 & \nodata \\ 
\nodata & 2005 Nov 24 & 53698.4 & 1.426 & \nodata & \nodata  & 25.27$\pm$0.84 & \nodata \\ 
\nodata & 2005 Nov 26 & 53700.4 & 1.427 & \nodata & \nodata  & 27.36$\pm$1.00 & \nodata \\ 
\nodata & 2005 Nov 27 & 53701.3 & 1.427 & \nodata & \nodata  & 26.83$\pm$1.02 & \nodata \\ 
\nodata & 2005 Dec 7 & 53711.3 & 1.432 & \nodata & \nodata  & 25.30$\pm$1.10 & \nodata \\ 
\nodata & 2006 Jan 11 & 53746.3 & 1.449 & \nodata & \nodata  & 24.54$\pm$0.83 & \nodata \\ 
\nodata & 2006 Jan 16 & 53751.2 & 1.452 & \nodata & \nodata  & 25.40$\pm$0.86 & \nodata \\ 
\nodata & 2006 Jan 19 & 53754.3 & 1.453 & \nodata & \nodata  & 24.79$\pm$0.92 & \nodata \\ 
\nodata & 2006 Jan 29 & 53764.2 & 1.458 & \nodata & \nodata  & 25.73$\pm$0.93 & \nodata \\ 
\nodata & 2006 Jan 31 & 53766.2 & 1.459 & \nodata & \nodata  & 26.24$\pm$0.96 & \nodata \\ 
\nodata &  2006 Feb 2 &	53768.2 &	    1.460 & 619.25$\pm$13.44  & \nodata & \nodata & \nodata \\
\nodata & 2006 Mar 14 & 53808.1 & 1.480 & \nodata & \nodata  & 25.60$\pm$0.81 & \nodata \\ 
\nodata &  2006 Mar 15 &	53809.2 &	    1.481 & 672.20$\pm$15.56 & \nodata & \nodata & \nodata  \\
\nodata & 2006 Mar 16 & 53810.1 & 1.481 & \nodata & \nodata  & 25.31$\pm$0.96 & \nodata \\ 
\nodata &  2006 Mar 19 &	53813.1 &	    1.482 & 662.69$\pm$16.13 & \nodata & \nodata & \nodata  \\
\nodata & 2006 Mar 20 & 53814.2 & 1.483 & \nodata & \nodata  & 24.44$\pm$1.22 & \nodata \\ 
\nodata & 2006 Apr 7 & 53832.1 & 1.492 & \nodata & \nodata  & 25.84$\pm$0.79 & \nodata \\ 
\nodata & 2006 Jun 3 & 53889.9 & 1.520 & \nodata & \nodata  & 26.65$\pm$0.68 & \nodata \\ 
\nodata & 2006 Aug 10 & 53958.0 & 1.554 & \nodata & \nodata  & 26.70$\pm$0.62 & \nodata \\ 
\nodata & 2006 Aug 12 & 53960.0 & 1.555 & \nodata & \nodata  & 25.93$\pm$0.95 & \nodata \\ 
\nodata & 2006 Oct 9 & 54017.4 & 1.583 & \nodata & \nodata  & 27.52$\pm$1.39 & \nodata \\ 
\nodata & 2006 Oct 11 & 54019.4 & 1.584 & \nodata & \nodata  & 27.13$\pm$1.28 & \nodata \\ 
\nodata &  2006 Oct 12 &	54020.4 &	    1.585 & 816.89$\pm$18.85 & \nodata & \nodata & \nodata  \\
\nodata & 2006 Oct 13 & 54021.4 & 1.585 & \nodata & \nodata  & 24.71$\pm$0.47 & \nodata \\ 
\nodata & 2006 Oct 16 & 54024.4 & 1.587 & \nodata & \nodata  & 28.00$\pm$0.96 & \nodata \\ 
\nodata & 2006 Dec 2 & 54071.3 & 1.610 & \nodata & \nodata  & 26.91$\pm$0.99 & \nodata \\ 
\nodata & 2006 Dec 4 & 54073.3 & 1.611 & \nodata & \nodata  & 26.76$\pm$0.89 & \nodata \\ 
\nodata & 2006 Dec 12 & 54081.4 & 1.615 & \nodata & \nodata  & 24.65$\pm$0.69 & \nodata \\ 
\nodata & 2006 Dec 14 & 54083.4 & 1.616 & \nodata & \nodata  & 23.61$\pm$0.43 & \nodata \\ 
\nodata & 2006 Dec 17 & 54086.2 & 1.618 & \nodata & \nodata  & 26.17$\pm$1.07 & \nodata \\ 
\nodata & 2006 Dec 21 & 54090.3 & 1.620 & \nodata & \nodata  & 26.53$\pm$1.03 & \nodata \\ 
\nodata & 2006 Dec 23 & 54092.3 & 1.621 & \nodata & \nodata  & 24.16$\pm$0.52 & \nodata \\ 
\nodata & 2007 Jan 18 & 54118.3 & 1.633 & \nodata & \nodata  & 27.30$\pm$1.28 & \nodata \\ 
\nodata & 2007 Jan 30 & 54130.4 & 1.639 & \nodata & \nodata  & 27.94$\pm$0.85 & \nodata \\ 
\nodata & 2007 Feb 1 & 54132.2 & 1.640 & \nodata & \nodata  & 26.92$\pm$1.51 & \nodata \\ 
\nodata & 2007 Feb 3 & 54134.3 & 1.641 & \nodata & \nodata  & 24.85$\pm$0.27 & \nodata \\ 
\nodata & 2007 Feb 5 & 54136.1 & 1.642 & \nodata & \nodata  & 27.89$\pm$1.09 & \nodata \\ 
\nodata & 2007 Feb 7 & 54138.1 & 1.643 & \nodata & \nodata  & 24.57$\pm$0.22 & \nodata \\ 
\nodata & 2007 Feb 9 & 54140.2 & 1.644 & \nodata & \nodata  & 27.20$\pm$1.15 & \nodata \\ 
\nodata & 2007 Feb 12 & 54143.2 & 1.646 & \nodata & \nodata  & 27.02$\pm$0.92 & \nodata \\ 
\nodata & 2007 Feb 14 & 54145.1 & 1.647 & \nodata & \nodata  & 27.20$\pm$0.76 & \nodata \\ 
\nodata & 2007 Feb 18 & 54149.1 & 1.649 & \nodata & \nodata  & 23.98$\pm$0.28 & \nodata \\ 
\nodata & 2007 Mar 31 & 54190.1 & 1.669 & \nodata & \nodata  & 26.51$\pm$0.74 & \nodata \\ 
\nodata &  2007 Apr 7 &	54197.1 &	    1.672 & 859.88$\pm$16.11  & \nodata & \nodata & \nodata \\
\nodata & 2007 Apr 12 & 54202.1 & 1.675 & \nodata & \nodata  & 26.46$\pm$0.94 & \nodata \\ 
\nodata &  2007 Apr 19 &	54209.2 &	    1.678 & 733.62$\pm$69.01 & \nodata & \nodata & \nodata  \\
\nodata & 2007 Jun 21 & 54272.9 & 1.710 & \nodata & \nodata  & 25.90$\pm$0.88 & \nodata \\ 
\nodata &  2007 Jun 26 &	54278.0 &	    1.712 & 779.16$\pm$16.16  & \nodata & \nodata & \nodata \\
\nodata & 2007 Jun 27 & 54279.0 & 1.713 & \nodata & \nodata  & 25.35$\pm$1.58 & \nodata \\ 
\nodata &  2007 Jun 29 &	54281.0 &	    1.714 & 838.81$\pm$24.54 & \nodata & \nodata & \nodata  \\
\nodata & 2007 Jul 2 &	54283.9 &    1.715 & 782.43$\pm$13.82  & \nodata & \nodata & \nodata \\
\nodata &  2007 Jul 10 &	54292.0 &	    1.719 & 809.56$\pm$16.48 & \nodata & \nodata & \nodata  \\
\nodata & 2007 Jul 18 & 54300.0 & 1.723 & \nodata & \nodata  & 25.75$\pm$1.03 & \nodata \\ 
\nodata & 2007 Jul 25 & 54306.0 & 1.726 & \nodata & \nodata  & 23.87$\pm$1.03 & \nodata \\ 
\nodata & 2007 Jul 28 & 54310.0 & 1.728 & \nodata & \nodata  & 26.84$\pm$1.11 & \nodata \\ 
\nodata & 2008 Feb 8 & 54504.3 & 1.824 & \nodata & \nodata  & 24.77$\pm$1.30 & \nodata \\ 
\nodata & 2008 Feb 13 & 54509.3 & 1.827 & \nodata & \nodata  & 25.06$\pm$0.98 & \nodata \\ 
\nodata & 2008 Feb 19 & 54515.3 & 1.830 & \nodata & \nodata  & 25.90$\pm$1.35 & \nodata \\ 
\nodata & 2008 Feb 26 & 54522.3 & 1.833 & \nodata & \nodata  & 25.98$\pm$1.00 & \nodata \\ 
\nodata & 2008 Feb 27 & 54523.2 & 1.833 & \nodata & \nodata  & 23.92$\pm$0.81 & \nodata \\ 
\nodata & 2008 Mar 1 & 54526.2 & 1.835 & \nodata & \nodata  & 25.57$\pm$1.31 & \nodata \\ 
\nodata & 2008 Mar 3 & 54528.2 & 1.836 & \nodata & \nodata  & 26.24$\pm$0.81 & \nodata \\ 
\nodata & 2008 Mar 7 & 54532.3 & 1.838 & \nodata & \nodata  & 25.06$\pm$1.14 & \nodata \\ 
\nodata & 2008 Mar 9 & 54534.2 & 1.839 & \nodata & \nodata  & 25.96$\pm$0.83 & \nodata \\ 
\nodata & 2008 Mar 15 & 54540.2 & 1.842 & \nodata & \nodata  & 27.94$\pm$0.81 & \nodata \\ 
\nodata & 2008 Mar 25 & 54550.2 & 1.847 & \nodata & \nodata  & 24.75$\pm$0.83 & \nodata \\ 
\nodata & 2008 Mar 30 & 54555.1 & 1.849 & \nodata & \nodata  & 28.42$\pm$1.06 & \nodata \\ 
\nodata & 2008 Apr 3 & 54559.1 & 1.851 & \nodata & \nodata  & 23.38$\pm$0.56 & \nodata \\ 
\nodata & 2008 Apr 14 & 54570.1 & 1.857 & \nodata & \nodata  & 27.13$\pm$1.09 & \nodata \\ 
\nodata & 2008 Apr 16 & 54572.1 & 1.858 & \nodata & \nodata  & 26.63$\pm$0.81 & \nodata \\ 
\nodata & 2008 Apr 22 & 54578.1 & 1.861 & \nodata & \nodata  & 23.87$\pm$0.59 & \nodata \\ 
\nodata & 2008 May 4 & 54590.1 & 1.867 & \nodata & \nodata  & 23.86$\pm$0.29 & \nodata \\ 
\nodata & 2008 May 11 & 54597.1 & 1.870 & \nodata & \nodata  & 24.89$\pm$0.59 & \nodata \\ 
\nodata & 2008 May 16 & 54602.1 & 1.873 & \nodata & \nodata  & 26.87$\pm$1.02 & \nodata \\ 
\nodata & 2008 Jun 22 & 54639.0 & 1.891 & \nodata & \nodata  & 25.03$\pm$0.61 & \nodata \\ 
\nodata & 2008 Jun 27 & 54645.0 & 1.894 & \nodata & \nodata  & 25.68$\pm$0.84 & \nodata \\ 
\nodata & 2008 Jul 3 & 54651.0 & 1.897 & \nodata & \nodata  & 23.89$\pm$0.67 & \nodata \\ 
\nodata & 2008 Jul 10 & 54657.0 & 1.900 & \nodata & \nodata  & 25.01$\pm$1.03 & \nodata \\ 
\nodata & 2008 Jul 13 & 54661.0 & 1.902 & \nodata & \nodata  & 24.60$\pm$0.50 & \nodata \\ 
\nodata & 2008 Nov 6 & 54776.3 & 1.959 & \nodata & \nodata  & 24.07$\pm$1.21 & \nodata \\ 
\nodata & 2008 Dec 4 & 54804.3 & 1.972 & \nodata & \nodata  & 22.46$\pm$0.92 & \nodata \\ 
\nodata & 2008 Dec 8 & 54808.3 & 1.974 & \nodata & \nodata  & 23.41$\pm$0.79 & \nodata \\ 
\nodata & 2008 Dec 15 & 54815.3 & 1.978 & \nodata & \nodata  & 22.97$\pm$1.34 & \nodata \\ 
\nodata & 2008 Dec 17 & 54817.3 & 1.979 & \nodata & \nodata  & 22.50$\pm$1.22 & \nodata \\ 
\nodata & 2008 Dec 22 & 54822.4 & 1.981 & \nodata & \nodata  & 20.97$\pm$1.10 & \nodata \\ 
\nodata & 2008 Dec 28 & 54828.4 & 1.984 & \nodata & \nodata  & 20.09$\pm$0.67 & \nodata \\ 
\nodata & 2009 Mar 8 & 54898.1 & 2.019 & \nodata & \nodata  & 16.21$\pm$0.02 & \nodata \\ 
\nodata & 2010 Jan 10 & 55206.3 & 2.171 & \nodata & \nodata  & 19.69$\pm$0.80 & \nodata \\ 
\nodata & 2010 Jan 21 & 55217.3 & 2.177 & \nodata & \nodata  & 20.75$\pm$0.79 & \nodata \\ 
\nodata & 2010 Apr 25 & 55312.0 & 2.223 & \nodata & \nodata  & 21.47$\pm$0.56 & \nodata \\ 
\nodata & 2010 Oct 30 & 	55499.38 & 2.316 &\nodata & \nodata  & 	20.81$\pm$1.14 & \nodata \\
\nodata & 2010 Nov 11 & 	55511.36 & 2.322 &\nodata & \nodata  & 	17.13$\pm$2.29 & \nodata \\
\nodata & 2010 Nov 16 & 	55516.3 & 2.324 &\nodata & \nodata  & 	20.76$\pm$1.23 & \nodata \\
\nodata & 2011 Mar 6 & 	55626.2 & 2.379 &\nodata & \nodata  & 	22.43$\pm$1.05 & \nodata \\
\nodata & 2011 Apr 17 & 	55668.1 & 2.399 &\nodata & \nodata  & 		20.76$\pm$1.10 & \nodata \\
\nodata & 2011 Nov 14 & 	55879.4 & 	2.504 &\nodata & \nodata  & 	21.23$\pm$1.06 & \nodata \\
\nodata & 2011 Nov 30 & 	55895.3 & 2.512 &\nodata & \nodata  & 		22.16$\pm$1.00 & \nodata \\
\nodata & 2011 Dec 16 & 	55911.3 & 2.520 &\nodata & \nodata  & 		20.68$\pm$0.91 & \nodata \\
\nodata & 2011 Dec 19 & 	55914.3 &	2.521 &\nodata & \nodata  & 	21.02$\pm$1.01 & \nodata \\
\nodata & 2011 Dec 21 & 	55916.2 & 2.522 &	\nodata & \nodata  & 	20.93$\pm$0.93 & \nodata \\
\nodata & 2011 Dec 23 & 55918.4 & 2.523 & 526.45$\pm$25.27 & \nodata	& \nodata & \nodata  \\
\nodata & 2011 Dec 28 & 	55923.2 & 2.526 &\nodata & \nodata  & 	21.64$\pm$1.01 & \nodata \\
\nodata & 2011 Dec 31 & 	55926.3 & 2.527 &\nodata & \nodata  & 		18.79$\pm$3.02 & \nodata \\
\nodata & 2012 Jan 5 & 	55931.2 & 2.530 &	\nodata & \nodata  & 	20.59$\pm$1.19 & \nodata \\
\nodata & 2012 Jan 17 & 	55943.4 & 2.536 &	\nodata & \nodata  & 	22.22$\pm$0.95 & \nodata \\

\enddata
	\tablecomments{}
	\tablenotetext{a}{As listed on the Eta Carinae Treasury Project site at \protect{http://etacar.umn.edu/}.}  
	\tablenotetext{b}{Integration between $\lambda\lambda$6520--6620, continuum at $\lambda\lambda$6500--6510 and $\lambda\lambda$6640--6645.}
		\tablenotetext{c}{Integration between $\lambda\lambda$4085--4115, continuum at $\lambda\lambda$4077--4078 and $\lambda\lambda$4155--4160.}
\end{deluxetable}

\end{document}